\pdfoutput=1
\documentclass[onecolumn]{sdl}

\def\0{\mbox{\tiny $0$}}
\def\1{\mbox{\tiny $1$}}
\def\2{\mbox{\tiny $2$}}
\def\3{\mbox{\tiny $3$}}
\def\4{\mbox{\tiny $4$}}
\def\5{\mbox{\tiny $5$}}
\def\6{\mbox{\tiny $6$}}
\def\7{\mbox{\tiny $7$}}
\def\8{\mbox{\tiny $8$}}
\def\9{\mbox{\tiny $9$}}

\logo{
\colorbox{DarkGoldenrod}{\color{white}$\mathbf{\Sigma\hspace*{0.06cm} \delta \hspace*{0.04cm} \Lambda}$}
}

\journal{\shadowtext{\textbf{\color{DarkRed} Annalen der Physik}}\,\,\, \textbf{529},\, 1600357-20 \, (2017).}

\titlelines{2}
\title{Optimizing weak measurements to \\ detect angular deviations}

\imgbgabstract{We analyze and compare the angular deviations for an optical beam reflected by and transmitted through a dielectric triangular prism. The analytic expressions derived for the angular deviations hold for arbitrary incidence angles. For incidence approaching the internal and external Brewster angles, the angular deviations transverse magnetic waves  present the same behavior leading to the well-known giant Goos-H\"anchen angular shift. For incidence near the critical angle a new region of large shift is seen both for transverse magnetic and transverse electric waves.  While a direct measuring procedure is better in the vicinity of the  Brewster region, a weak measurement breaks off the giant Goos-H\"anchen effect, preserving the amplification in the critical region. We discuss under which conditions it is  possible to optimize the amplification and we also determine when a weak measurement is preferred to a direct measuring procedure.}

\author{
\names{Manoel P. Ara\'ujo$^{1,a}$, Stefano De Leo$^{2,b}$ and Gabriel G. Maia$^{1,c}$}
\affiliation{$^{1}$Institute of Physics Gleb Wataghin, State University of Campinas, Brazil.}
\email{$^{a}$mparaujo@ifi.unicamp.br $\,\,\,\,\, $ $^{c}$ggm11@ifi.unicamp.br.}
\affiliation{$^{2}$Department of Applied Mathematics, State of University of Campinas, Brazil.}
\email{$^{b}$deleo@ime.unicamp.br}
}

\begin{document}

\sdlmaketitle

\section{Introduction}

It is well known that finite angular distributions of optical beams introduce corrections to the path of light predicted by geometrical optics\cite{born,saleh}. For instance, lateral and transverse displacements of laser beams, when totally reflected by an internal face of an optical prism, are explained by the Goos-H\"{a}nchen \cite{GH1,GH2,GH3,GH4,GH5,GH6,GH7,GH8,GH9,GH10} and the Imbert-Fedorov \cite{GHIF0,GHIF1,GHIF2,GHIF3,GHIF4} effects, respectively. {\color{black} The breaking of symmetry of the optical beam induces angular deviations {\cite{ANG1,ANG2,ANG3,ANG4,ANG5,ANG6,ExpANG1,ExpANG2}}. The Fresnel filtering \cite{FF,ExpFF1} can be interpreted as a manifestation of such deviations.} In the critical region, as it has been shown recently \cite{AGHE,Break,Axial}, {\color{black} angular deviations mix with the lateral displacement leading} to a composite shift \cite{CGH}. {\color{black}The Fresnel filtering phenomenon owns its name to the role the Fresnel coefficients play in the reflection of an optical beam by an air/dielectric or dielectric/air interface.} When a plane wave hits the interface between two dielectric media, part of it is reflected with the same angle of incidence and part is refracted according to the Snell law. The amount of light reflected by and transmitted through the second medium is determined by the Fresnel coefficients. A plane wave has a well defined wave vector and, consequently, it has a unique direction of incidence. This is not the case for an incident Gaussian beam with a finite angular distribution centred around the incidence angle $\theta_{\0}$,
\begin{equation}
\label{gauss}
g(\theta-\theta_{\0}) \, = \, \frac{k\,{\mathrm{w}}_{\0}}{2\,\sqrt{\pi}} \, \exp\left[-(\theta-\theta_{\0})^{^2}(k \,{\mathrm{w}}_{\0})^{^2}/\,4\right]\,\,,
\end{equation}
where $k=2\,\pi/\lambda$, $\lambda$ is the beam wavelength, and ${\mathrm{w}}_{\0}$ is the beam waist. The finite angular distribution generates a range of incoming wave vectors. If,  for every angle in the angular distribution, the Fresnel coefficients contribute with the same weight, there is no preferred transmission of one angle over the others and the geometrical laws still hold. If, however, the coefficients differ from angle to angle, this difference will filter the beam,  breaking its symmetry. Angular deviations from geometrical optics predictions are the net result of this filtering process and have also been confirmed  by several optical experiments\cite{ExpANG1,ExpANG2,ExpFF1}.

In the interval in which, { for a given incident angle $\theta_{\0}$}, the angular distribution of a Gaussian laser beam, Eq.\,(\ref{gauss}), is significant,
\[\theta_{\0} \,-\,  \displaystyle{\frac{\lambda}{{\mathrm{w}}_{\0}}}\, <\,\,\theta\,\,< \,\theta_{\0}\, +\,  \displaystyle{\frac{\lambda}{{\mathrm{w}}_{\0}}}\,\,, \]
the Fresnel coefficients are, almost everywhere, a very smooth function of the incidence angle, the plane wave limit holds, and the angular deviations are very small, {being proportional to $1/(k\,{\mathrm{w}}_{\0})^{\2}$}. Nevertheless, for incidence near the Brewster angles, the Fresnel coefficients of Transverse Magnetic (TM) waves break the symmetry of the angular distribution, { increasing the angular deviation, making it} proportional to $1/k \,{\mathrm{w}}_{\0}$, generating what is known in the literature as giant Goos-H\"anchen effect \cite{GiantGH}. This is caused by the change of sign in the TM Fresnel coefficients when the incidence angle is the Brewster angle. Since Transverse Electric (TE) waves have positive Fresnel coefficients for every incidence angle, this effect is not present for such waves.

{ Something similar can be observed near the critical angle. In this region the Fresnel reflection coefficient becomes complex and its absolute value changes abruptly, generating a noticeable break in the beam's symmetry}. 

{ Angular deviations have been studied for a long time now. In 1973 Ra \emph{et al.} \cite{ANG1} identified the phenomenon by a direct analysis of the integrated reflected beam expression for the case of internal reflections. In their work they recognize the critical angle as a threshold between lateral and angular shifts but do not find the deviation expression at it nor do they analyze the Brewster angle. One year later Antar and Boerner \cite{ANG2}, studying a similar system, found an expression for the angular deviation at the Brewster angle. They express the reflected beam as a summation of reflected beam modes where the fundamental mode is a Gaussian beam. At the Brewster angle this fundamental mode is absent, being then the angular deviation due to higher order modes. They also identified a change in the sign of the angular shift relatively to the direction predicted by geometrical optics as the incident angle moves from a value smaller than the Brewster angle to a value greater than it. In 1977, White \emph{et al.} \cite{ANG3} gave a more physical interpretation of the phenomenon, in contrast to the more mathematical approaches of the earlier studies. They introduced the interpretation of the angular shift as the result of a change in the power distribution of the plane wave components of the beam. Studying the reflection of a light beam coming from a denser medium into a rarer one, they approximated the propagation direction of the reflected beam as the propagation direction of the plane wave which, in the far field, had the biggest contribution to the power of the reflected spectrum. By doing so, they found a double value for the angular deviation at the Brewster angle because, contrary to the method of Antar and Boerner, which considers the shift of the whole double peaked object (so the angular deviation is the mean deviation), White \emph{et al.} considered the angular shift of each peak individually, recognizing that these shifts were slightly different because the Fresnel reflection coefficient is not symmetric around the Brewster angle. In 1985, Chan and Tamir \cite{ANG4} analyzed the problem of the angular shift at and around the Brewster angle also by a direct analysis of the integrated reflected beam expression. In their paper they argue that the double peak structure is so distorted in relation to the incident Gaussian that the concept of angular shifts is meaningless. This distortion effect at the Brewster angle have been more carefully studied more recently by Aiello et al. \cite{CrossPol}. One year later, in 1986, Tamir and Chan, in a remarkable review of all beam phenomena occurring near critical incidence \cite{ANG5}, they reproduced Ra \emph{et al.} formula and also calculated the deviation value for incidence exactly at the critical angle. In 2009, Aiello and Woerdman \cite{ANG6} revisited the angular deviations near the Brewster angle for an air/glass interface, calculating the beam shift as the mean distance between the beam center and the axis of propagation predicted by geometrical optics. Parallel to these developments, known as the angular Goos-H\"{a}nchen shift, the microcavity community studied an angular deviation of their own, mostly associated to transmission instead of reflection, known as Fresnel filtering. This term and this field of research was inaugurated by Tureci and Stone in 2002 \cite{FF}. In their paper the authors point out that critical incidence on a dielectric/air interface does not lead to a transmission tangent to the boundary, instead, very large deviations from this expectation occur. Such deviations have been a matter of interest in the microcavities community as can be seen in \cite{MicCav}. Tureci and Stone presented an explanation for this effect similar to the one given by White \emph{et al.}, being the similarity of the phenomena confirmed theoretically in 2013 by G\"otte \emph{et al.} \cite{ANGFF}. It is curious that, while experimental results in microcavities \cite{ExpFF1} prompted the theoretical study of Tureci and Stone, more than thirty years passed between the theoretical prediction of the angular Goos-H\"anchen shift and its experimental confirmation. Due to its minute nature it was firstly observed in 2006 in the microwave domain, by M\"uller \emph{et al.} \cite{ExpANG1}, being measured in the optical
domain three years later by Merano \emph{et al.} \cite{ExpANG2}.}

{ The small magnitude of the angular Goos-H\"anchen shift suggests the employment of amplification techniques in its detection. In 1988, Aharonov, Albert, and Vaidman (AAV) presented their seminal paper on the amplification technique they called \emph{weak measurement} (WM) \cite{WM1}. Their work was developed in a quantum mechanical framework and was based on a weak interaction between a system and a meter, that is, another system used to measure some property of the first one. Their idea was that by a careful choice of final states a weak interaction between system and meter could provide a trade off between the probability of obtaining the state and the eigenvalue it would return. By paying the price of a very low probability event, the measured value associated to it (in their original work they discuss spin measurements) could be greatly increased. The details of their set-up are beyond the scope of our paper, but the interested reader can refer to the excellent review of the subject provided in ref. \cite{WM2}. In 1989, Duck and Stevenson published a paper addressing a few formal inconsistencies of AAV's manuscript, but acknowledging their results \cite{WM3}. In the same paper they also presented an optical analogue of the weak measurement. In this classical perspective the trade off is between the intensity of a light beam and an induced deviation of its path. Both quantum and optical weak measurements have been successfully employed in amplifying small signal phenomena. The optical case is a well established tool, particularly in Goos-H\"{a}nchen shifts measurements \cite{ExpWM1}, but it has also been used for angular deviations, especially for air/dielectric reflections \cite{ExpWM2,ExpWM3}.

The purpose of this work is to consider how efficient this technique is in comparison to a direct measurement. To do so}, we consider a Gaussian laser beam interacting with a dielectric right angle triangular prism of refractive index $n$ as depicted in Fig.\,1.(a). This optical system has the advantage of allowing us to, with a single set-up, study several regions of interest. When the beam hits the left face of the prism part of it is reflected and part transmitted. The reflected beam (REF) reproduces the case commonly studied in angular deviation weak measurements, with the possibility to study the external Brewster angles. The transmitted beam will propagate until it hits the lower interface, being then, once again, divided in a transmitted and a reflected beam. The beam transmitted through the lower interface (LTRA) allows the study of the critical angle while for the beam reflected (and then finally transmitted through the right interface, which we call an upper transmission (UTRA)) one internal Brewster angle and the critical angle are available. This upper transmitted beam provides a good region for comparisons between Brewster and critical angular shifts and is in fact the region we are most interested in. To help in our description of this system five coordinate systems are employed, see Fig. \,1(b). The left coordinate system, $y-z$, has the $z-$axis perpendicular to the left face of the prism and the $y-$axis parallel to it. The lower coordinate system, $y_{_{\ast}}-z_{_{\ast}}$, has the $z_{_{\ast}}-$axis perpendicular to the lower interface of the prism and the $y_{_{\ast}}-$axis parallel to it. These are the prism coordinate systems. The other three are beam coordinate systems. They follow the propagation direction predicted by the geometrical optics for the reflected, $y_{_{\mathrm{REF}}}-z_{_{\mathrm{REF}}}$, the lower transmitted, $y_{_{\mathrm{LTRA}}}-z_{_{\mathrm{LTRA}}}$, and the upper transmitted, $y_{_{\mathrm{UTRA}}}-z_{_{\mathrm {UTRA}}}$ beams. Their propagation direction is parallel to their respective $z-$axes, while the $y-$axes are the transversal axes.

{ The paper is structured as follows. In section II we discuss the external and internal Brewster regions and the critical region as well. These regions are the range of incidence angles around the Brewster and critical angles which have the beam's structure modified by the proximity to these particular values. The regions' frontiers are determined by waist the of the beam. Section III is devoted to the calculation of the electric field amplitudes of the incident beam and of all beams resultant from its interaction with the prism and of their associated powers. The incidence angle $\theta_{\0}$ will be considered the independent variable for all cases, since it is the variable the experimentalist has actual control over. In section IV, by a calculation of the mean optical path, analytic formulas are given for the angular deviations taking place at all faces of the prism and approximations are given for incidence in the proximity of the Brewster and critical angles. We show that the external and internal Brewster regions provide the same angular deviation as each other, which is independent of the refractive index of the dielectric, while for incidence close to the critical region, the angular shift is refractive index dependent. We also compare in this section the angular shifts in the Brewster region to the ones in the critical region, the two regions with the largest values for this sort of shift. In section V we present the weak measurement experimental set-up. Our results show that while the weak measurement technique is very powerful near the critical region its amplification efficiency is greatly diminished near the Brewster region. We discuss how to optimize this amplification. Conclusions are drawn in the final section.}

\section{The Brewster and critical angles}

{For external reflections, {\color{black} the reflection at the left side of the prism, see Fig.1(a)}, it is possible to divide the range of incident angles into five subregions, established around the two Brewster angles existent for this sort of reflection,}
\[
\begin{array}{r}
 {\color{black}{\theta_{\0}^{^{\mathrm{[I]}}}}} < \theta_{_{\mathrm{B(ext)}}}^{^{-}} - \,\displaystyle{\frac{\lambda}{\,{\mathrm{w}}_{\0}}} < \,\, {\color{black}{\theta_{\0}^{^{\mathrm {[II]}}}}} <\,
 \theta_{_{\mathrm{B(ext)}}}^{^{-}} + \, \displaystyle{\frac{\lambda}{\,{\mathrm{w}}_{\0}}}\\
 \theta_{_{\mathrm{B(ext)}}}^{^{-}} + \, \displaystyle{\frac{\lambda}{\,{\mathrm{w}}_{\0}}}    < \,\, {\color{black}{\theta_{\0}^{^{\mathrm {[III]}}}}} <\,\theta_{_{\mathrm{B(ext)}}}^{^{+}} - \,\displaystyle{\frac{\lambda}{\,{\mathrm{w}}_{\0}}}\\
\theta_{_{\mathrm{B(ext)}}}^{^{+}} - \,\displaystyle{\frac{\lambda}{\,{\mathrm{w}}_{\0}}}
 <  \,\,{\color{black}{\theta_{\0}^{^{\mathrm{[IV]}}}}} <
 \theta_{_{\mathrm{B(ext)}}}^{^{+}} + \,\displaystyle{\frac{\lambda}{\,{\mathrm{w}}_{\0}}}  <
 \,\, \theta_{\0}^{^{\mathrm{[V]}}} \\
 \end{array}
\]
{where $\theta_{_{\mathrm{B(ext)}}}^{\pm}$ are the Brewster angles, given by}
\begin{equation}{\label{eq:BE}}
\theta_{_{\mathrm{B(ext)}}}^{^{\pm}}=\pm \arcsin\left[\,n\,/\sqrt{n^{^{2}}+1}\,\right]\,\,,
\end{equation}
{and whose derivation will be discussed in the next section.} For incidence angles in the region I, III, and V the Gaussian angular distribution for the reflected TM waves does not meet the Brewster region and its shape is not drastically changed as it happens for incidence in the region II and IV where the giant Goos-H\"anchen effect takes place. The reader interested in a more detailed discussion is recommended to look over the very interesting paper cited in ref. \cite{ANG4}.

The reflection at the down dielectric/air interface, Fig.1.(a), creates, in addition to the internal Brewster region, a \emph{new} breaking of symmetry region near the critical angle. {\color{black} In terms of the internal reflection angle $\varphi_{\0}$, we find the following five subregions:
\[
\begin{array}{r}
 \varphi_{\0}^{^{\mathrm{[I]}}} < \varphi_{_{\mathrm{B(int)}}} - \,\displaystyle{\frac{\lambda}{\,{\mathrm{w}}_{\0}}} < \,\, \varphi_{\0}^{^{\mathrm  {[II]}}} <\,
 \varphi_{_{\mathrm{B(int)}}} + \, \displaystyle{\frac{\lambda}{\,{\mathrm{w}}_{\0}}}\\
 \varphi_{_{\mathrm{B(int)}}} + \, \displaystyle{\frac{\lambda}{\,{\mathrm{w}}_{\0}}}    < \,\, \varphi_{\0}^{^{\mathrm{[III]}}} <\,\varphi_{_{\mathrm {cri}}} - \,\displaystyle{\frac{\lambda}{\,{\mathrm{w}}_{\0}}}\\
\varphi_{_{\mathrm{cri}}} - \,\displaystyle{\frac{\lambda}{\,{\mathrm{w}}_{\0}}}
 <  \,\,\varphi_{\0}^{^{\mathrm{[IV]}}} <
 \varphi_{_{\mathrm{cri}}} + \,\displaystyle{\frac{\lambda}{\,{\mathrm{w}}_{\0}}}  <
 \,\, \varphi_{\0}^{^{\mathrm{[V]}}} \\
 \end{array}
\]
where the internal Brewster angle is given by
\begin{equation}
\varphi_{_{\mathrm{B(int)}}}= \arcsin\left[\,1\,/\sqrt{n^{^{2}}+1}\,\right]\,\,,
\end{equation}
and the critical angle by
\begin{equation}
\varphi_{_{\mathrm{cri}}} = \arcsin[1/n].
\end{equation}
By using the geometry of the system, $\varphi_{\0}=\psi_{\0}+\pi/4$, and the Snell law, $\sin\theta_{\0}=n\,\sin\psi_{\0}$, see Fig.\,1(a), we can divide the incidence region for internal reflection in the following subregions,}

\[
\begin{array}{r}
 {\color{black}{\theta_{\0}^{^{\mathrm{[I]}}}}} < \theta_{_{\mathrm{B(int)}}} - \,\displaystyle{\frac{\lambda}{\,{\mathrm{w}}_{\0}}} < \,\, {\color{black}{\theta_{\0}^{^{\mathrm{[II]}}}}} <\,
 \theta_{_{\mathrm{B(int)}}} + \, \displaystyle{\frac{\lambda}{\,{\mathrm{w}}_{\0}}}\\
 \theta_{_{\mathrm{B(int)}}} + \, \displaystyle{\frac{\lambda}{\,{\mathrm{w}}_{\0}}}    < \,\, {\color{black}{\theta_{\0}^{^{\mathrm{[III]}}}}} <\,\theta_{_{\mathrm{cri}}} - \,\displaystyle{\frac{\lambda}{\,{\mathrm{w}}_{\0}}}\\
\theta_{_{\mathrm{cri}}} - \,\displaystyle{\frac{\lambda}{\,{\mathrm{w}}_{\0}}}
 <  \,\,{\color{black}{\theta_{\0}^{^{\mathrm{[IV]}}}}} <
 \theta_{_{\mathrm{cri}}} + \,\displaystyle{\frac{\lambda}{\,{\mathrm{w}}_{\0}}}  <
 \,\, {\color{black}{\theta_{\0}^{^{\mathrm{[V]}}}}} \\
 \end{array}
\]
where
\begin{equation}
\label{BCRI}
\begin{array}{l}
\theta_{_{\mathrm{B(int)}}}=\arcsin\left[\,n\,(1-n)\,/\sqrt{2\,(n^{^{2}}+1)}\,\right]\,\,,\\
\theta_{_{\mathrm{cri}}}=\arcsin\left[\,\left(\,1- \sqrt{n^{^{2}}-1}\,\right)\,/\,\sqrt{2}\,\right]\,\,.
\end{array}
\end{equation}
{\color{black}We take the liberty to introduce these critical and internal Brewster incidence angles for two reasons. The first one is to differentiate them from the standard Brewster angles, used in describing  external reflection, and the second is because these are the angles directly accessible for experimentalists in the laboratory}. It is important to observe that, differently from the external reflection case, the internal Brewster angle, as well the critical angle, depend on the geometrical characteristics  of the optical prism. The angles given in Eq.\,(\ref{BCRI}) refer to a right angle triangular prism whose planar section is represented in Fig.\,1.(a)

The importance to study the internal reflection is clearly seen by observing that we have the possibility not only to reproduce the giant Goos-H\"anchen effect as in the external reflection case but also to study, in proximity of the critical angle, a new amplification region.

{For transmissions through the lower dielectric/air interface of the prism the range of incidence angles can be divided into two subregions around the critical angle. In terms of the internal incidence angle we have that
\[
\varphi_{\0}^{^{\mathrm{[I]}}} < \varphi_{_{\mathrm{cri}}} - \,\displaystyle{\frac{\lambda}{\,{\mathrm{w}}_{\0}}} < \,\, \varphi_{\0}^{^{\mathrm{[II]}}} <\,
 \varphi_{_{\mathrm{cri}}} + \, \displaystyle{\frac{\lambda}{\,{\mathrm{w}}_{\0}}},
\]
where
\[\varphi_{_{\mathrm{cri}}} = \arcsin[1/n].\]
By using the geometry of the system, see Fig.\,1(a), these regions can be expressed for an incidence angle $\theta_{\0}$ at the left face of the prism,
\[
\theta_{\0}^{^{\mathrm{[I]}}} < \theta_{_{\mathrm{cri}}} - \,\displaystyle{\frac{\lambda}{\,{\mathrm{w}}_{\0}}} < \,\, \theta_{\0}^{^{\mathrm{[II]}}} <\,
 \theta_{_{\mathrm{cri}}} + \, \displaystyle{\frac{\lambda}{\,{\mathrm{w}}_{\0}}},
\]

Region I is a well defined angular deviation region. In region II part of the beam incident upon the lower interface is already undergoing total internal reflection and so the transmitted beam begins to become an evanescent field. Such fields are out of the scope of our paper and shall not be considered.}

\section{Incident, reflected, and transmitted beams}

As incident beam, let us consider a free Gaussian laser modelled by the angular  distribution given in Eq.\,(\ref{gauss}).
The incidence angle  $\theta_{\0}$ is the angle that the beam forms with the normal  to the left (air/dielectric) interface, the $z-$axis in Fig.\,1(b).  The amplitude of the incident {\color{black}(INC)} electric field is expressed by

\begin{equation}
E_{_{\mathrm{INC}}}\, = \,   E_{\0}\,\int_{_{-\pi/2}}^{^{+\pi/2}} \hspace*{-0.5cm} {\mathrm{d}}\theta \,\,\, g(\theta-\theta_{\0}) \,\, \exp[\,i \, k \, (\,y\,\sin\theta+z\,\cos\theta\,)\,]\,\,.
\end{equation}

In the paraxial approximation, $\lambda\ll {\mathrm{w}}_{_{0}}$, we can expand the spatial phase up to the second order around the incidence angle $\theta_{_{0}}$, obtaining

\begin{eqnarray*}
y\,\sin\theta_{_{0}}+z\,\cos\theta_{_{0}} + (\,y\,\cos\theta_{_{0}}-z\,\sin\theta_{_{0}}\,)\,(\theta-\theta_{_{0}}) -  (\,y\,\sin\theta_{_{0}}+z\,\cos\theta_{_{0}}\,)\,(\theta-\theta_{_{0}}\,)^{^{2}}/2\,\,.
\end{eqnarray*}
Using this second order expansion and the incident axes, see Fig.\,1(b),
\begin{eqnarray}
\left[ \begin{array}{cc} y_{_{\mathrm{INC}}} \\ z_{_{\mathrm{INC}}} \end{array} \right] \, = \, \left( \begin{array}{rrrr} \cos\theta_{\0} & -\,\sin\theta_{\0} \\ \sin\theta_{\0} & \cos\theta_{\0} \end{array} \right) \, \left[ \begin{array}{cc} y\\ z \end{array} \right]\,\,,
\end{eqnarray}

{\noindent}we can rewrite the amplitude of the incident beam as follows
\begin{equation}{\label{eq:Einc}}
\begin{array}{r}
E_{_{\mathrm{INC}}}\, = \, E_{\0}\,\displaystyle{\int_{_{-\pi/2}}^{^{+\pi/2}}} \hspace*{-0.5cm} {\mathrm{d}}\theta \,\,\, g(\theta - \theta_{\0}) \,\exp\{\,i \, k \, [\,y_{_{\mathrm{INC}}} \, (\theta - \theta_{\0}) - z_{_{\mathrm{INC}}} \, (\theta - \theta_{\0})^{^{2}}/2\,]\,\}\,\,.
\end{array}
\end{equation}
The previous expression represents a Gaussian beam with its minimal beam waist at the point in which the beam touches the left (air/dielectric) interface, see Fig.1(a). This has its mathematical counterpart in choosing the origin of our axes in such a point, see Fig.\,1.

{ The incident power is given by
\begin{equation}{\label{eq:Pinc}}
\begin{array}{r}
P_{_{\mathrm{INC}}}\, = \, \displaystyle{\int_{_{-\infty}}^{^{+\infty}}} \hspace*{-0.35cm} {\mathrm{d}}y_{_{\mathrm{INC}}} \,\,\, \left|E_{_{\mathrm{INC}}}\right|^{^2}.
\end{array}
\end{equation}
Using the relation
\begin{equation}{\label{eq:diracdelta}}
\begin{array}{r}
\displaystyle{\int_{_{-\infty}}^{^{+\infty}}} \hspace*{-0.4cm} {\mathrm{d}}
y_{_{\mathrm{INC}}}\,\,e^{{i\,k\,(\,\theta-\widetilde{\theta}\,)\,y_{_{\mathrm{INC}}}}} =\,\,\frac{2\,\pi}{k}\,\delta(\,\theta-\widetilde{\theta}\,),
\end{array}
\end{equation}
where $\delta(\theta-\tilde{\theta})$ is the Dirac delta function, we obtain for Eq. (\ref{eq:Pinc})
\begin{equation}{\label{eq:Pinc2}}
P_{_{\mathrm{INC}}} \, = \, \frac{2\,\pi}{k}\,\left|E_{\0}\right|^{^2}\,\int_{_{-\infty}}^{^{+\infty}}\hspace*{-0.5cm} {\mathrm{d}}\theta \,\,\, g^{\2}(\theta - \theta_{\0})\,=\, \sqrt{\frac{\pi}{2}}\,{\mathrm{w}}_{\0}\,\left|E_{\0}\right|^{^2}.
\end{equation}}

\subsection{ The reflected beam}

The beam reflected (REF) at the left interface of the prism has its angular spectrum modified by the reflection coefficient of the interface,
\begin{equation}
\begin{array}{r}
E_{_{\mathrm{REF}}}^{^{\mathrm{[TE,TM]}}}\, = \,  E_{\0}\, \displaystyle{ \int_{_{-\pi/2}}^{^{+\pi/2}}} \hspace*{-0.5cm} {\mathrm{d}}\theta \,\,\, g(\theta-\theta_{\0}) \,\,R_{_{\mathrm{left}}}^{^{\mathrm{[TE,TM]}}}(\theta)\, \exp[\,i \, k \, (\,y\,\sin\theta-z\,\cos\theta\,)\,]\,\,,
\end{array}
\end{equation}
where
\begin{equation}
\left\{\begin{array}{l}R_{_{\mathrm{left}}}^{^{\mathrm{[TE]}}}(\theta) \\ \\ R_{_{\mathrm{left}}}^{^{\mathrm{[TM]}}}(\theta) \end{array}\right\}\, = \,
\left\{\,
\begin{array}{l}
\displaystyle{\frac{ \cos \theta - n\,\cos \psi}{\cos\theta + n\,\cos\psi}} \\ \\
\displaystyle{\frac{n  \cos \theta - \cos \psi}{n\,\cos\theta + \cos\psi}}
\end{array}
\right\},
\end{equation}
being $\psi$ given by the Snell law, $\sin\theta=n\,\sin\psi$. As done for the incident beam, we develop the spatial phase up to the second order around $\theta_{\0}$,
\begin{eqnarray*}
y\,\sin\theta_{_{0}}-z\,\cos\theta_{_{0}} + (\,y\,\cos\theta_{_{0}}+z\,\sin\theta_{_{0}}\,)\,(\theta-\theta_{_{0}}) - (\,y\,\sin\theta_{_{0}}-z\,\cos\theta_{_{0}}\,)\,(\theta-\theta_{_{0}}\,)^{^{2}}/2\,\,,
\end{eqnarray*}
and using the reflected axes, see Fig.\,1(b),
\begin{equation}
\begin{array}{l}
\left[ \begin{array}{cc} z_{_{\mathrm{REF}}} \\ y_{_{\mathrm{REF}}} \end{array} \right] \, = \,
\left( \begin{array}{rrrr} \sin\theta_{\0} & -\,\cos\theta_{\0} \\ \cos\theta_{\0} & \sin\theta_{\0} \end{array} \right) \, \left[ \begin{array}{cc} y\\ z \end{array} \right]\,\,,
\end{array}
\end{equation}
we obtain for the reflected beam the following expression
\begin{equation}
\begin{array}{r}
E_{_{\mathrm{REF}}}^{^{\mathrm{[TE,TM]}}}\, = \, E_{\0}\,\displaystyle{\int_{_{-\pi/2}}^{^{+\pi/2}}} \hspace*{-0.5cm} {\mathrm{d}}\theta \,\,\, g(\theta - \theta_{\0}) \,\,R_{_{\mathrm{left}}}^{^{\mathrm{[TE,TM]}}}(\theta)\,\exp\{\,i \, k \, [\,y_{_{\mathrm{REF}}} \, (\theta - \theta_{\0}) - z_{_{\mathrm{REF}}} \, (\theta - \theta_{\0})^{^{2}}/2\,]\,\}\,\,.
\end{array}
\end{equation}
{ The reflected power calculation is similar to the one performed in Eq. (\ref{eq:Pinc2}), with the difference that the Gaussian function is multiplied by the reflection coefficient
\begin{equation}
P_{_{\mathrm{REF}}} \, = \, \frac{2\,\pi}{k}\,\left|E_{\0}\right|^{^2}\,\int_{_{-\infty}}^{^{+\infty}}\hspace*{-0.5cm} {\mathrm{d}}\theta \,\,\, \left[g(\theta - \theta_{\0})R_{_{\mathrm{left}}}^{^{\mathrm{[TE,TM]}}}(\theta)\right]^{^{2}}.
\end{equation}
Expanding $\left[R_{_{\mathrm{left}}}^{^{\mathrm{[TE,TM]}}}(\theta)\right]^{\2}$ up to first order we can integrate the equation above, obtaining the normalized reflected power as
\begin{equation}
\begin{array}{r}
\mathcal{P}_{_{\mathrm{REF}}} \, = \, \frac{\displaystyle P_{_{\mathrm{REF}}}}{\displaystyle P_{_{\mathrm{INC}}}} \, = \, \left|R_{_{\mathrm{left}}}^{^{\mathrm{[TE,TM]}}}(\theta_{\0})\right|^{^2}.
\end{array}
\end{equation}
Notice that, for $n\,\cos\theta_{\0} = \cos\psi_{\0}$, $R_{_{\mathrm{left}}}^{^{\mathrm{[TM]}}}(\theta_{\0}) = 0$. This happens for the Brewster angles in Eq. (\ref{eq:BE}),
\[\theta_{_{\mathrm{B(ext)}}}^{^{\pm}}=\pm\arcsin\left[\,n\,/\,\sqrt{n^{\2}+1}\,\right]\,\,,   \]
which are, for this reason, also called polarization angles. For a BK7 ($n=1.515$) prism, they are $\pm 56.573^{\circ}$.

The normalized reflected power for TE and TM polarized beams as a function of the incidence angle $\theta_{\0}$ is plotted in Fig.2(a).}

\subsection{The upper transmitted beam}

The upper transmission (UTRA), which follows an internal reflection, will have the angular spectrum of beam modified by the transmission coefficients of the left and right interfaces and by the reflection coefficient of the lower interface as well,
\begin{eqnarray}
E_{_{\mathrm{UTRA}}}^{^{\mathrm{[TE,TM]}}}  =  E_{\0}\,\displaystyle{\int_{_{-\pi/2}}^{^{+\pi/2}}} \hspace*{-0.5cm} {\mathrm{d}}\theta \,g(\theta-\theta_{\0})
\, T_{_{\mathrm{left}}}^{^{\mathrm{[TE,TM]}}}(\theta)\,R_{_{\mathrm{down}}}^{^{\mathrm{[TE,TM]}}}(\theta)\,
T_{_{\mathrm{right}}}^{^{\mathrm{[TE,TM]}}}(\theta){\nonumber}\\
\times\,\exp[i\,\Phi_{_{\mathrm{UGEO}}}(\theta)]\,\exp[i \, k \, (\,z\,\sin\theta+y\,\cos\theta\,)\,]\,\,,
\end{eqnarray}
where
\begin{equation}
\left\{\begin{array}{l}T_{_{\mathrm{left}}}^{^{\mathrm{[TE]}}}(\theta)\,T_{_{\mathrm{right}}}^{^{\mathrm{[TE]}}}(\theta) \\ \\ T_{_{\mathrm{left}}}^{^{\mathrm{[TM]}}}(\theta)\,T_{_{\mathrm{right}}}^{^{\mathrm{[TM]}}}(\theta) \end{array}\right\}
= \left\{\,\begin{array}{l}
 \displaystyle{  \frac{4 \, n \, \cos\theta \, \cos\psi}{(\cos\theta + n \, \cos\psi)^{^2}}}\\ \\
 \displaystyle{  \frac{4 \, n \, \cos\theta \, \cos\psi \, }{(n \, \cos\theta+\cos\psi)^{^2}}}
\end{array}
\, \right\}
\end{equation}
are the product of the transmission coefficients and
\begin{equation}
\left\{\begin{array}{l}R_{_{\mathrm{down}}}^{^{\mathrm{[TE]}}}(\theta) \\ \\ R_{_{\mathrm{down}}}^{^{\mathrm{[TM]}}}(\theta \end{array}\right\} =
\left\{\,\begin{array}{l}
 \displaystyle{ \frac{n \, \cos\varphi - \cos\phi}{n \, \cos\varphi + \cos\phi}}\\ \\
 \displaystyle{ \frac{\cos\varphi-n \, \cos\phi}{\cos\varphi + n \, \cos\phi}}
\end{array}
\, \right\}
\end{equation}
the reflection coefficients of the lower interface. The choice to put the origin of our axes at the point in which the beam touches the left (air/dielectric) interface, which implies a minimal beam waist for the incident and reflected beam at the proximity of the left interface, requires to use in addition to the standard Fresnel coefficients for the transmission through the left (air/dielectric) and right (dielectric/air) interfaces and for the reflection at the down (dielectric/air) interface, the geometrical phase\cite{geophs},
\begin{eqnarray}
 \Phi_{_{\mathrm{UGEO}}}(\theta) = k\,\left[ \, (\,\cos\theta - \sin\theta\,)\,a
 +\, \left(\,n\,\cos\psi\,-\,\cos\theta\,\right)\,b \, \right]\,\,,
\end{eqnarray}
where $a$ is the distance between the origin of the axes and the left down corner of the triangular planar section of the prism and $b$ is the value of the shortest sides of the planar section, see Fig.\,1(a).

Expanding the spatial phase up to the second order around the incidence angle,
\begin{eqnarray*}
z\,\sin\theta_{_{0}}+y\,\cos\theta_{_{0}} + \\ (\,z\,\cos\theta_{_{0}}-y\,\sin\theta_{_{0}}\,)\,(\theta-\theta_{_{0}}) - \\ (\,z\,\sin\theta_{_{0}}+y\,\cos\theta_{_{0}}\,)\,(\theta-\theta_{_{0}}\,)^{^{2}}/2\,\,,
\end{eqnarray*}
and using the upper transmitted axes, see Fig.\,1(b),
\begin{equation}
\left[ \begin{array}{cc} z_{_{\mathrm{UTRA}}} \\ y_{_{\mathrm{UTRA}}} \end{array} \right] \, = \, \left( \begin{array}{rrrr} \cos\theta_{\0} & \sin\theta_{\0} \\ -\,\sin\theta_{\0} & \cos\theta_{\0} \end{array} \right) \, \left[ \begin{array}{cc} y\\ z \end{array} \right]\,\,,
\end{equation}
we obtain
{
\begin{eqnarray}
E_{_{\mathrm{UTRA}}}^{^{\mathrm{[TE,TM]}}} = E_{\0} \,\displaystyle{\int_{_{-\pi/2}}^{^{+\pi/2}}} \hspace*{-0.5cm} {\mathrm{d}}\theta \,\,\, g(\theta-\theta_{\0})\,
T_{_{\mathrm{left}}}^{^{\mathrm{[TE,TM]}}}(\theta)\,R_{_{\mathrm{down}}}^{^{\mathrm{[TE,TM]}}}(\theta)\,
T_{_{\mathrm{right}}}^{^{\mathrm{[TE,TM]}}}(\theta)\\
\times\,\exp[i\,\Phi_{_{\mathrm{UGEO}}}(\theta)]\,
\exp\{i \, k \, [\,y_{_{\mathrm{UTRA}}} \, (\theta - \theta_{\0}) - z_{_{\mathrm{UTRA}}} \, (\theta - \theta_{\0})^{^{2}}/2\,]\,\}\,\,.
\end{eqnarray}
}
Expanding, as done for the spatial phase, the geometrical phase up to the second order around $\theta_{\0}$, we can further simplify the expression of the transmitted beam
{
\begin{eqnarray}
\!\!E_{_{\mathrm{UTRA}}}^{^{\mathrm{[TE,TM]}}} =  E_{\0} \, \displaystyle{\int_{_{-\pi/2}}^{^{+\pi/2}}} \hspace*{-0.5cm} {\mathrm{d}}\theta \,\,\, g(\theta-\theta_{\0})
T^{^{\mathrm{[TE,TM]}}}_{_{\mathrm{left}}}(\theta)R_{_{\mathrm{down}}}^{^{\mathrm{[TE,TM]}}}(\theta)T^{^{\mathrm{[TE,TM]}}}_{_{\mathrm{right}}}(\theta)\\
\times\,\exp\{\,i \, k \, [\,\widetilde{y}_{_{\mathrm{UTRA}}} \, (\theta - \theta_{\0}) - \widetilde{z}_{_{\mathrm{UTRA}}} \, (\theta - \theta_{\0})^{^{2}}/2\,]\,\}\,\,,
\end{eqnarray}
}
where
\[
\begin{array}{l}
\widetilde{y}_{_{\mathrm{UTRA}}}=y_{_{\mathrm{UTRA}}}+\, \Phi^{^{\prime}}_{_{\mathrm{GEO}}}(\theta_{\0})\,/\,k=y_{_{\mathrm{UTRA}}} - y_{_{\mathrm{GEO}}}\,\,,\\ \\
\widetilde{z}_{_{\mathrm{UTRA}}}=z_{_{\mathrm{UTRA}}}- \,\Phi^{^{\prime\prime}}_{_{\mathrm{GEO}}}(\theta_{\0})\,/\,k=z_{_{\mathrm{UTRA}}} - z_{_{\mathrm{GEO}}}\,\,.
\end{array}
\]
The first derivative of the geometrical phase calculated at $\theta_{\0}$ and divided by $k$,
\begin{eqnarray*}
y_{_{\mathrm{UGEO}}}= (\,\sin\theta_{\0} + \cos\theta_{\0}\,)\,\,a\,+\,
  \left(\,\frac{\cos\theta_{\0}}{n\,\cos\psi_{\0}}\,-\,1\,\right)\,\sin\theta_{\0}\,\,b\,\,.
\end{eqnarray*}
gives the exit point of the beam along the $y_{_{\mathrm{LTRA}}}$ axis. The use of the stationary phase method to obtain the beam shift represents an alternative way, with respect to the one based on the Snell law, to calculate  the geometrical path of the optical beam \cite{geophs}. While the first derivative reproduces the optical path predicted by geometrical optics, it is interesting to observe that the second derivative acts as a beam profile modifier. This was theoretically suggested in ref.\cite{profTeo} and recently experimentally confirmed \cite{profExp}. The geometrical shift of the transmitted beam has the effect of centering it around $y_{_{\mathrm{LGEO}}}$ and, as the beam profile modification does not interfere in the calculation of the mean value of the transversal component, the second phase derivative does not affect the calculation of the angular deviations.

Following the procedure done for the normalized reflected power the normalized upper transmitted power is given by
\begin{equation}
\mathcal{P}_{_{\mathrm{UTRA}}} = \left| T^{^{\mathrm{[TE,TM]}}}_{_{\mathrm{left}}}(\theta_{\0})R_{_{\mathrm{down}}}^{^{\mathrm{[TE,TM]}}}(\theta_{\0})T^{^{\mathrm{[TE,TM]}}}_{_{\mathrm{right}}}(\theta_{\0})\right|^{^2}.
\end{equation}

For the TE and TM transmitted beam, we find  total internal reflection for  incidence greater than the critical one
\begin{eqnarray*}
\begin{array}{cc}\,n\,\sin\varphi_{_{\mathrm{cri}}}= 1 \\ \\ \Rightarrow \,\theta_{_{\mathrm{cri}}}= \arcsin\left[\,\left(\,1- \sqrt{n^{^2}-1}\,\,\right)\,/\,\sqrt{2}\,\right].\,\,\end{array}
\end{eqnarray*}
For a BK7 prism, this incidence angle is $\theta_{_{\mathrm{cri}}}=-5.603^{^{o}}$, see Fig.\,2(b). For the TM beam, we also find  a  Brewster angle at $n\,\sin\varphi_{_{\mathrm{B(int)}}}= \,\cos\varphi_{_{\mathrm{B(int)}}}$ which implies \[\theta_{_{\mathrm{B(int)}}}= \arcsin\left[\,(\,1 - n\,)\,n\,/\,\sqrt{\,2\,(n^{\2}+1)}\,\right]\,\,.   \]
For a BK7 prism, the Brewster angle for internal reflection is found at $\theta_{_{\mathrm{B(int)}}}=-17.693^{^{o}}$, see Fig.\,2(b).

The relative upper transmitted power for TE and TM polarization is plotted against the incidence angle $\theta_{\0}$ in Fig.2(b).

\subsection{The lower transmitted beam}

{The beam transmitted through the lower interface (LTRA) will have its Gaussian profile modified by two Fresnel coefficients, the transmission coefficients of the left and the lower interfaces,
\begin{eqnarray}
E_{_{\mathrm{LTRA}}}^{^{\mathrm{[TE,TM]}}} =  E_{\0}\, \displaystyle{ \int_{_{-\pi/2}}^{^{+\pi/2}}} \hspace*{-0.5cm} {\mathrm{d}}\theta \,\,\, g(\theta-\theta_{\0})
T_{_{\mathrm{left}}}^{^{\mathrm{[TE,TM]}}}(\theta)\,T_{_{\mathrm{down}}}^{^{\mathrm{[TE,TM]}}}(\theta)\\
\times \, \exp[\,i\,\Phi_{_{\mathrm{LGEO}}}(\theta)\,]
\exp[\,i \, k \, (\,y_{_\ast}\,\sin\phi+z_{_\ast}\,\cos\phi\,)\,]\,\,,
\end{eqnarray}
where
\begin{equation}
\left\{\, \begin{array}{l}
T_{_{\mathrm{left}}}^{^{\mathrm{[TE]}}}(\theta)\\  \\
T_{_{\mathrm{left}}}^{^{\mathrm{[TM]}}}(\theta)
\end{array}
\,\right\}
= \left\{\,\begin{array}{l}
 \displaystyle{  \frac{2 \, \cos\theta}{\cos\theta + n\,\cos\psi}}\\ \\
 \displaystyle{  \frac{2 \,n\, \cos\theta}{n\,\cos\theta+ \cos\psi}}
\end{array}
\, \right\}
\end{equation}
are the transmission coefficients of the left interface and
\begin{equation}
\left\{\, \begin{array}{l}
T_{_{\mathrm{down}}}^{^{\mathrm{[TE]}}}(\theta)\\  \\
T_{_{\mathrm{down}}}^{^{\mathrm{[TM]}}}(\theta)
\end{array}
\,\right\}
= \left\{\,\begin{array}{l}
 \displaystyle{  \frac{2 \, n \, \cos\varphi}{n\,\cos\varphi + \cos\phi}}\\ \\
 \displaystyle{  \frac{2 \, \cos\varphi}{\cos\varphi+ n\,\cos\phi}}
\end{array}
\, \right\},
\end{equation}
are the lower transmission coefficients. $\phi$ is the transmission angle, see Fig\,1(a), obtained from $n\,\sin\varphi = \sin\phi$.  $\Phi_{_{\mathrm {LGEO}}}(\theta)$ is the geometrical phase of the lower transmitted beam. It is given by
\begin{equation}
\begin{array}{l}
 {\displaystyle\Phi_{_{\mathrm{LGEO}}}(\theta) = \frac{k\,a}{\sqrt{2}}\,(n\,\cos\varphi-\cos\phi)}\,\,.
 \end{array}
 \end{equation}

Here, we can once again expand the beam's spatial phase up to second order and then write it in the lower transmission coordinate system, but the expansion will be different than the one carried out for the reflected and transmitted beams because it is a function of $\theta$ through $\phi$. It will be

\begin{eqnarray*}
y_{_\ast}\,\sin\phi_{_{0}}+z_{_\ast}\,\cos\phi_{_{0}} + \\ (\,y_{_\ast}\,\cos\phi_{_{0}}-z_{_\ast}\,\sin\phi_{_{0}}\,)\,\phi^{\prime}_{\0}\,(\theta-\theta_{_{0}})\,-\\   (\,y_{_\ast}\,\sin\phi_{_{0}}-z_{_\ast}\,\cos\phi_{_{0}}\,)\,(\phi^{\prime}_{\0})^{\2}\,(\theta-\theta_{_{0}}\,)^{^{2}}/2 + \\
(\,y_{_\ast}\,\sin\phi_{_{0}}+z_{_\ast}\,\cos\phi_{_{0}}\,)\,\phi^{\prime\prime}_{\0}\,(\theta-\theta_{_{0}}\,)^{^{2}}/2
\,\,,
\end{eqnarray*}

{\noindent}where
\begin{equation}{\label{eq:phiprime}}
\phi^{\prime}_{\0}\, = \, \frac{\displaystyle \cos\varphi_{\0}\,\cos\theta_{\0}}{\displaystyle \cos\psi_{\0}\,\cos\phi_{\0}}
\end{equation}
is the first order derivative of $\phi$, with respect to $\theta$, evaluated at $\theta_{\0}$, and $\phi_{\0}^{\prime\prime}$ is the second order derivative.

Using the lower transmitted axis, see Fig.1(b),
\begin{equation}
\left[ \begin{array}{cc} y_{_{\mathrm{LTRA}}} \\ z_{_{\mathrm{LTRA}}} \end{array} \right] \, = \, \left( \begin{array}{rrrr} \cos\phi_{\0} & -\sin\phi_{\0} \\ \sin\phi_{\0} & \cos\phi_{\0} \end{array} \right) \, \left[ \begin{array}{cc} y_{_\ast}\\ z_{_\ast} \end{array} \right]\,\,,
\end{equation}
we obtain the electric field of the beam transmitted through the lower interface as
\begin{eqnarray}
E_{_{\mathrm{LTRA}}}^{^{\mathrm{[TE,TM]}}}\, = \, E_{\0}\,\displaystyle{\int_{_{-\pi/2}}^{^{+\pi/2}}} \hspace*{-0.5cm} {\mathrm{d}}\theta \,\,\, g(\theta - \theta_{\0})\,T_{_{\mathrm{left}}}^{^{\mathrm{[TE,TM]}}}(\theta)\,T_{_{\mathrm{down}}}^{^{\mathrm{[TE,TM]}}}(\theta)\\
\exp[\,i\,\Phi_{_{\mathrm{LGEO}}}(\theta)]\,
\times\,\exp\{\,i \, k \, (\,y_{_{\mathrm{LTRA}}} \, \phi_{\0}^{\prime}\,(\theta - \theta_{\0}) - z_{_{\mathrm{LTRA}}} \,[\phi_{\0}^{\prime}\,(\theta - \theta_{\0})]^{^{2}}/2\,)\,\}.
\end{eqnarray}
Notice that we have not considered in the spatial phase expansion the $\phi^{\prime\prime}_{\0}$ term. This is justified because $\phi^{\prime}_{\0}\sim\phi^{\prime\prime}_{\0}$ and $z_{_{\mathrm{LTRA}}} \gg y_{_{\mathrm{LTRA}}}$, meaning that measurements are carried out at a distance far greater than the beam's width and so we can neglect the term containing this second order derivative. Expanding now, as done for the spatial phase, the lower geometrical phase up to the second order around $\theta_{\0}$, we can simplify the expression of the transmitted beam to
\begin{equation}
\begin{array}{l}
E_{_{\mathrm{LTRA}}}^{^{\mathrm{[TE,TM]}}}\, = \,  E_{\0} \, \displaystyle{\int_{_{-\pi/2}}^{^{+\pi/2}}} \hspace*{-0.5cm} {\mathrm{d}}\theta \,\,\, g(\theta-\theta_{\0})\, T^{^{\mathrm{[TE,TM]}}}_{_{\mathrm{left}}}(\theta)\,T^{^{\mathrm{[TE,TM]}}}_{_{\mathrm{down}}}(\theta)\,\, \times\\
\exp\{\,i \, k \, [\,\widetilde{y}_{_{\mathrm{LTRA}}} \, (\theta - \theta_{\0}) - \widetilde{z}_{_{\mathrm{LTRA}}} \, (\theta - \theta_{\0})^{^{2}}/2\,]\,\}\,\,,
\end{array}
\end{equation}
where
\[
\begin{array}{l}
\widetilde{y}_{_{\mathrm{LTRA}}}=\phi^{\prime}_{\0}\,y_{_{\mathrm{LTRA}}}+\, \Phi^{^{\prime}}_{_{\mathrm{LGEO}}}(\theta_{\0})\,/\,k=\phi^{\prime}_{\0}\,y_{_{\mathrm{LTRA}}}- y_{_{\mathrm{LGEO}}}\,\,,\\ \\
\widetilde{z}_{_{\mathrm{LTRA}}}=(\phi^{\prime}_{\0})^{\2}\,z_{_{\mathrm{LTRA}}}- \,\Phi^{^{\prime\prime}}_{_{\mathrm{LGEO}}}(\theta_{\0})\,/\,k = (\phi^{\prime}_{\0})^{\2}\,z_{_{\mathrm{LTRA}}}-z_{_{\mathrm{LGEO}}}\,\,.
\end{array}
\]
The first derivative of the geometrical phase calculated at $\theta_{\0}$ and divided by $k$ is
\begin{equation*}
y_{_{\mathrm{LGEO}}} = \frac{a}{\sqrt{2}}\,\left(\,\tan\varphi_{\0}\,\cos\phi_{\0}-\sin\phi_{\0}\,\right)\,\,.
\end{equation*}\,

The normalized lower transmitted power, following what was done for the reflected power is
\begin{equation}{\label{eq:LPwr}}
\begin{array}{r}
\mathcal{P}_{_{\mathrm{LTRA}}}\, = \, \frac{\displaystyle 1}{\displaystyle \phi^{\prime}_{\0}}\left|T_{_{\mathrm{left}}}^{^{\mathrm{[TE,TM]}}}(\theta_{\0})\,T_{_{\mathrm{down}}}^{^{\mathrm{[TE,TM]}}}(\theta_{\0})\right|^{^2}.
\end{array}
\end{equation}
Notice that for the lower transmitted power, the integral in Eq. (\ref{eq:diracdelta}) will have a factor $\phi^{\prime}_{\0}$ multiplying $y_{_{\mathrm {LTRA}}}$ and so the Dirac delta will have a factor $1/\phi^{\prime}_{\0}$ multiplying it. This justifies this factor in Eq. (\ref{eq:LPwr}). It is interesting to note that lower transmitted power is proportional to $\cos\phi_{\0}$, see Eq.(\ref{eq:phiprime}), and so it goes to zero as the incident angle approaches $\theta_{_{\mathrm{cri}}}$. 
The normalized transmitted power is plotted for TE and TM polarization against the incidence angle $\theta_{\0}$ in Fig.2(c).}

\section{Analytical expressions for the angular deviations}

The decisive steps in obtaining an analytic expression for the angular deviations are the use of the mean value calculation of the transversal component and the observation that the spatial integration can be converted into an angular integration. Here, the symmetry of the angular distribution plays a fundamental role. For a free Gaussian beam, same as the one  propagating from the source to the dielectric block, we have to calculate

\begin{equation}
\langle\, y_{_{\mathrm{INC}}} \rangle =  \int_{_{-\infty}}^{^{+\infty}} \hspace*{-0.4cm} {\mathrm{d}}
y_{_{\mathrm{INC}}}\,\, y_{_{\mathrm{INC}}} \,\, |E_{_{\mathrm{INC}}}|^{^2}\,\,\,\mbox{\huge /}\,\,
 \int_{_{-\infty}}^{^{+\infty}} \hspace*{-0.4cm} {\mathrm{d}}
y_{_{\mathrm{INC}}}\,\, |E_{_{\mathrm{INC}}}|^{^2}\,\,.
\end{equation}

By using Eq. (\ref{eq:diracdelta}) and
{\[
\begin{array}{l}
\displaystyle{\int_{_{-\infty}}^{^{+\infty}}} \hspace*{-0.4cm} {\mathrm{d}}
y_{_{\mathrm{INC}}}\,\,y_{_{\mathrm{INC}}}\,\,e^{{i\,k\,(\,\theta-\widetilde{\theta}\,)\,y_{_{\mathrm{INC}}}}} =\,\,\frac{i\,\pi}{k^{\2}}\,[\,\partial_{_{\widetilde{\theta}}}\,\delta(\,\theta-\widetilde{\theta}\,)\,-\, \partial_{_{\theta}}\,\delta(\,\theta-\widetilde{\theta}  \,)]\,\,,
  \end{array}
  \]}
and applying an integration by parts,  the spatial integration can be transformed into its angular counterpart,
\begin{eqnarray}
\langle\, y_{_{\mathrm{INC}}} \rangle & = &\frac{\displaystyle{
  \int_{_{-\pi/2}}^{^{+\pi/2}} \hspace*{-0.4cm} {\mathrm{d}}
\theta\,\, (\theta-\theta_{_{0}})\, g^{\2}(\theta-\theta_{_{0}}) }}{\displaystyle{\int_{_{-\pi/2}}^{^{+\pi/2}} \hspace*{-0.4cm} {\mathrm{d}}
\theta\,\,g^{\2}(\theta-\theta_{_{0}}) }}\,\,\,z_{_{\mathrm{INC}}}\,\,.
\end{eqnarray}
In the paraxial approximation, $k{\mathrm{w}}_{\0}\gg 1$, the angular distribution is strictly peaked around the incidence angle $\theta_{\0}$ and, without loss of generality,  for incidence between $-\pi/2 +\lambda/{\mathrm{w}}_{\0}$ and $\pi/2 -\lambda/{\mathrm{w}}_{\0}$,  the limits of integration can be extended to $\pm \,\infty$. The symmetry of the Gaussian angular distribution implies a null value for the integral which appears in the numerator and, consequently, the beam  propagates along the $z_{_{\mathrm{INC}}}$ axis, i.e. $\langle\, y_{_{\mathrm{INC}}} \rangle =0$.

\subsection{Reflected beam}

Repeating for the {\color{black}external} reflected beam the steps  carried out for the incident one, we find

\begin{eqnarray}
\label{meanref}
\langle\, y_{_{\mathrm{REF}}}^{^{\mathrm{[TE,TM]}}} \rangle & = &\frac{\displaystyle{
  \int_{_{-\pi/2}}^{^{+\pi/2}} \hspace*{-0.4cm} {\mathrm{d}}
\theta\,\, (\theta-\theta_{_{0}})\, \left[\,g(\theta-\theta_{_{0}})\,R_{_{\mathrm{left}}}^{^{{\mathrm{[TE,TM]}}}}(\theta)\,\right]^{^{2}}}}{\displaystyle{\int_{_{-\pi/2}}^{^{+\pi/2}} \hspace*{-0.4cm} {\mathrm{d}}
\theta\,\,\left[\, g(\theta-\theta_{_{0}})\,R_{_{\mathrm{left}}}^{^{{\mathrm{[TE,TM]}}}}(\theta)\,\right]^{^{2}} }}\,\,\,z_{_{\mathrm{REF}}}.
\end{eqnarray}
In order to obtain an analytic solution, we can develop the reflection coefficient up to the second order around the incidence angle $\theta_{\0}$ to obtain
\[
\left[\,\frac{R_{_{\mathrm{left}}}^{^{{\mathrm{[TE,TM]}}}}(\theta)}{R_{_{\mathrm{left}}}^{^{{\mathrm{[TE,TM]}}}}(\theta_{\0})}\,\right]^{^{2}}
 =
 1\, +\,\,2\,\, \frac{R_{_{\mathrm{left}}}^{^{^{{\mathrm{[TE,TM]}}^\prime}}}(\theta_{\0})}{R_{_{\mathrm{left}}}^{^{^{\mathrm{[TE,TM]}}}}(\theta_{\0})}\,\,(\theta-\theta_{\0})\, \,+\]
 \[
  \left\{
  \,\left[\, \frac{R_{_{\mathrm{left}}}^{^{^{{\mathrm{[TE,TM]}}^\prime}}}(\theta_{\0})}{R_{_{\mathrm{left}}}^{^{^{\mathrm{[TE,TM]}}}}(\theta_{\0})}\,\right]^{^{2}} \,+\,\,
\frac{R_{_{\mathrm{left}}}^{^{^{{\mathrm{[TE,TM]}}^{\prime\prime}}}}(\theta_{\0})}{R_{_{\mathrm{left}}}^{^{^{\mathrm{[TE,TM]}}}}(\theta_{\0})}\,\right\}\,\,(\theta-\theta_{\0})^{^{2}}.
\]

Due to the symmetric nature of $g(\theta-\theta_{\0})$, in the numerator of Eq.(\ref{meanref}), the only contribution comes from the first order term expansion while in the denominator the contributions come from the zeroth and second order terms. In the second order term, due to the fact that the denominator of the first addend goes to zero more rapidly than the denominator of the second one, we can neglect it. Finally, without loss of generality, we can use the following approximation

  \begin{equation}
  \begin{array}{l}
 \displaystyle{  \left[\,\frac{R_{_{\mathrm{left}}}^{^{{\mathrm{[TE,TM]}}}}(\theta)}{R_{_{\mathrm{left}}}^{^{{\mathrm{[TE,TM]}}}}(\theta_{\0})}\,\right]^{^{2}}}
 =
 1\, +\,\,2\,\,D_{_{\mathrm{REF}}}^{^{{\mathrm{[TE,TM]}}}}(\theta_{\0})\,\,(\theta-\theta_{\0})\, +\\
 \left[\,D_{_{\mathrm{REF}}}^{^{{\mathrm{[TE,TM]}}}}(\theta_{\0})\,\right]^{^{2}} \,(\theta-\theta_{\0})^{^{2}}
\end{array}
\end{equation}

{\noindent}where
\begin{eqnarray}
D_{_{\mathrm{REF}}}^{^{{\mathrm{[TE,TM]}}}}(\theta_{\0}) &=& R_{_{\mathrm{left}}}^{^{^{{\mathrm{[TE,TM]}}^\prime}}}(\theta_{\0})\,\,/\,
R_{_{\mathrm{left}}}^{^{^{\mathrm{[TE,TM]}}}}(\theta_{\0})\,\,{\nonumber}\\
&=&\frac{2\,\sin\theta_{\0}}{n\,\cos\psi_{\0}}\,\left\{\, 1 \,,\, \frac{n^{\2}}{\sin^{^2}\theta_{\0} - n^{^{2}}\cos^{^{2}}\theta_{\0}}  \,\right\}\,\,.
\end{eqnarray}
This expansion and the identity
\[ \int_{_{-\pi/2}}^{^{+\pi/2}} \hspace*{-0.4cm} {\mathrm{d}}
\theta\,\, (\theta-\theta_{_{0}})^{^{2}}\,\,g^{\2}(\theta-\theta_{_{0}})\,=\, \frac{1}{\,\,\,(\,k\,{\mathrm{w}}_{\0}\,)^{^{2}}}\,\,\int_{_{-\pi/2}}^{^{+\pi/2}} \hspace*{-0.4cm} {\mathrm{d}}
\theta\,\,g^{\2}(\theta-\theta_{_{0}})\,\,,
\]
allow us to analytically solve the integrals appearing in  Eq.(\ref{meanref}). The angular deviations for the reflected beam are then given by
{
\begin{equation}
\label{adr}
\langle\, y_{_{\mathrm{REF}}}^{^{\mathrm{[TE,TM]}}} \rangle  =  \alpha_{_{\mathrm{REF}}}^{^{\mathrm{[TE,TM]}}}(\theta_{\0})\,\,\,z_{_{\mathrm{REF}}}
\end{equation}
where
\begin{equation*}
 \alpha_{_{\mathrm{REF}}}^{^{\mathrm{[TE,TM]}}}(\theta_{\0})= \displaystyle{   \frac{2\, D_{_{\mathrm{REF}}}^{^{{\mathrm{[TE,TM]}}}}(\theta_{\0})}{(k\,{\mathrm{w}}_{\0})^{^{^2}}\,+\, \left[\,D_{_{\mathrm{REF}}}^{^{{\mathrm{[TE,TM]}}}}(\theta_{\0})\,\right]^{^{2}}}}\,\,\,z_{_{\mathrm{REF}}}\,\,.
 \end{equation*}
 }
For incidence at the Brewster angles, observing that
\[  \left\{\,
D_{_{\mathrm{REF}}}^{^{{\mathrm{[TE]}}}}[\theta_{_{\mathrm{B(ext)}}}^{^{\pm}}]
\,,\,D_{_{\mathrm{REF}}}^{^{{\mathrm{[TM]}}}}[\theta_{_{\mathrm{B(ext)}}}^{^{\pm}}]
\, \right\}\,\,\to\,\,\left\{\, \pm\,\,2/n \,,\,\infty\,\right\}\,\,,
 \]
 we find
 \begin{equation}
 \left\{\,
 \alpha_{_{\mathrm{REF}}}^{^{\mathrm{[TE]}}}[\theta_{_{\mathrm{B(ext)}}}^{^{\pm}}]\,,\,
 \alpha_{_{\mathrm{REF}}}^{^{\mathrm{[TM]}}}[\theta_{_{\mathrm{B(ext)}}}^{^{\pm}}]
 \,\right\}\,=\,
 \left\{\,
 \pm\,\frac{4/n}{\,\,\,(\,k\,{\mathrm{w}}_{\0}\,)^{^{2}}}\,,\,
 0\,\right\}\,\,.
\end{equation}
It is interesting to calculate the angular deviations for TM waves for incidence in the close vicinity of the Brewster angle. A way to analyze the Fresnel coefficients behavior in the Brewster region consists in introducing a new parameter $\delta$ in the incidence angle,
\[ \theta_{\0}= \theta_{_{\mathrm{B(ext)}}}^{^{\pm}} \,+\,\, \frac{\delta}{k\,{\mathrm{w}}_{\0}}\,\,. \]
Observing that
\begin{eqnarray*}
\sin^{^2}\theta_{\0} - n^{^{2}}\cos^{^{2}}\theta_{\0} \,&\approx &
2\,\sin  \theta_{_{\mathrm{B(ext)}}}^{^{\pm}}
\,[\, \cos \theta_{_{\mathrm{B(ext)}}}^{^{\pm}}  \pm n\,\sin  \theta_{_{\mathrm{B(ext)}}}^{^{\pm}}  \,]\,\,\displaystyle{\frac{\delta}{k\,{\mathrm{w}}_{\0}}}\,\\
&=&\pm\,\,2\,n\,\, \frac{\displaystyle \delta}{\displaystyle k\,{ \mathrm{w}}_{\0}}\,\,,
\end{eqnarray*}
we have
\[
D_{_{\mathrm{REF}}}^{^{{\mathrm{[TM]}}}}\left[\theta_{_{\mathrm{B(ext)}}}^{^{\pm}} \,+\,\,\displaystyle{\frac{\delta}{k\,{\mathrm{w}}_{\0}}}\right]
 \,= \,\frac{k\,{\mathrm{w}}_{\0}}{\delta}
\]
which implies
\begin{equation}
\alpha_{_{\mathrm{REF}}}^{^{\mathrm{[TM]}}} \left[\theta_{_{\mathrm{B(ext)}}}^{^{\pm}} \,+\,\,\displaystyle{\frac{\delta}{k\,{\mathrm{w}}_{\0}}}\right] \,=\,\frac{2\,\delta}{1+ \delta^{^{2}}}\,\frac{1}{k\,{\mathrm{w}}_{\0}}\,\,.
\end{equation}
As a result of the previous calculation, we find that the angular deviations in the proximity of the Brewster angles do not depend on the refractive index. A simple $\delta$ derivation also shows that the maximal angular deviations are found at $\delta=\pm\,1$. For incidence in the Brewster region, we have for TM waves a $k\,{\mathrm{w}}_{\0}$ relative factor of the angular deviation with respect to the TE case. This is responsible for what is known in the literature as the giant Goos-H\"anchen angular shift\,\cite{GiantGH}. { Analytic expression similar to Eq.\,(\ref{adr}) have recently been obtained for an air/glass plane interface by A. Aiello and Woerdman in the paper of ref.\,\cite{ANG6}. In that paper, the authors calculated in an alternative way both the Goos-H\"anchen and Imbert-Federov spatial and angular shifts. For the Goos-H\"anchen angular deviations, they observed, in the Brewster region, the same behaviour plotted in Fig.\,3(a).}

\subsection{The upper transmitted beam}

Obviously, the considerations done in the previous subsection can be immediately repeated for and extended to the beam {\color{black}transmitted, after internal reflection, through the right side of the prism,} leading to
\begin{equation}
\label{meantra}
\langle\, y_{_{\mathrm{UTRA}}}^{^{\mathrm{[TE,TM]}}} \rangle
  = y_{_{\mathrm{UGEO}}} + \alpha_{_{\mathrm{UTRA}}}^{^{\mathrm{[TE,TM]}}}(\theta_{\0})\,(z_{_{\mathrm{UTRA}}}-z_{_{\mathrm{UGEO}}})
\end{equation}
where
\begin{equation*}
\alpha_{_{\mathrm{UTRA}}}^{^{\mathrm{[TE,TM]}}}(\theta_{\0})\,=\,
\frac{2\, D_{_{\mathrm{UTRA}}}^{^{{\mathrm{[TE,TM]}}}}(\theta_{\0})}{(k\,{\mathrm{w}}_{\0})^{^{^2}}\,+\, [\,D_{_{\mathrm{UTRA}}}^{^{{\mathrm{[TE,TM]}}}}(\theta_{\0})\,]^{^{^{^2}}}}
\end{equation*}
and
\begin{eqnarray}
D_{_{\mathrm{UTRA}}}^{^{{\mathrm{[TE,TM]}}}} &=&  R_{_{\mathrm{down}}}^{^{^{{\mathrm{[TE,TM]}}^\prime}}}(\theta_{\0})\,\,/\,
R_{_{\mathrm{down}}}^{^{^{\mathrm{[TE,TM]}}}}(\theta_{\0}) {\nonumber}\\
&=&\frac{2\,\sin\varphi_{\0}\,\cos\theta_{\0}}{\cos\phi_{\0}\,\cos\psi_{\0}}
\displaystyle{ \left\{\, 1 \,,\, \frac{1}{n^{\2}\sin^{\2}\varphi_{\0} - \cos^{^{2}}\varphi_{\0}}  \,\right\}}\,\,.{\nonumber}\\
\end{eqnarray}
Note that $T_{_{\mathrm{left}}}^{^{^{\mathrm{[TE,TM]}}}}$ and $T_{_{\mathrm{right}}}^{^{^{\mathrm{[TE,TM]}}}}$ are smooth functions and consequently the main contribution to angular deviations comes from $R_{_{\mathrm{down}}}^{^{^{\mathrm{[TE,TM]}}}}$.

The first analysis at the Brewster angle,
\[
\begin{array}{l}
\displaystyle{
\left\{\,
D_{_{\mathrm{UTRA}}}^{^{{\mathrm{[TE]}}}}[\theta_{_{\mathrm{B(int)}}}]
\,,\,
D_{_{\mathrm{UTRA}}}^{^{{\mathrm{[TM]}}}}[\theta_{_{\mathrm{B(int)}}}]
\, \right\}\,\,\to\,\,
\frac{2\,\cos\theta_{_{\mathrm{B(int)}}} }{\cos\psi_{_{\mathrm{B(int)}}}}\,
\left\{\, 1 \, \,,\,\infty\,\right\}
\,\,=\,\,}\\
\displaystyle{\left\{\, \frac{2\,\sqrt{n^{\2}-n^{\4}+2+2\,n^{\3}}}{n+1}\, \,,\,\infty\,\right\}},
 \end{array}
 \]
gives
 \begin{equation}
 \left\{\,
 \alpha_{_{\mathrm{UTRA}}}^{^{\mathrm{[TE]}}}[\theta_{_{\mathrm{B(int)}}}]\,,\,
 \alpha_{_{\mathrm{UTRA}}}^{^{\mathrm{[TM]}}}[\theta_{_{\mathrm{B(int)}}}]
 \,\right\}\,=\,
 \left\{\, \frac{4\,\sqrt{n^{\2}-n^{\4}+2+2\,n^{\3}}}{(n+1)\,\,(\,k\,{\mathrm{w}}_{\0}\,)^{^{2}}}
 \,,\,
 0\,\right\}.
\end{equation}
A more accurate study for the TM waves in the Brewster region can be carried out by introducing the parameter $\delta$ in the incident angle,
\[
\theta_{\0}= \theta_{_{\mathrm B(int)}} \,+\,\, \frac{\delta}{k\,{\mathrm{w}}_{\0}}\,\,\,\,\,\,\,\Rightarrow\,\,\,\,\,\,\,\,
\varphi_{\0}= \varphi_{_{\mathrm{B(int)}}} \,+\,\, \frac{\cos\theta_{_{\mathrm{B(int)}}}}{n\,\cos\psi_{_{\mathrm{B(int)}}}}\,\,\frac{\delta}{k\,{\mathrm{w}}_{\0}}\,\,.
 \]
 Observing that
 \begin{eqnarray*}
  n\,^{^2} \sin^{^2}\varphi_{\0} - \cos^{^{2}}\varphi_{\0}  & \approx & 2\,n\,\sin  \varphi_{_{\mathrm{B(int)}}}
\,[\, n\,\cos \varphi_{_{\mathrm{B(int)}}} + \, \sin  \theta_{_{\mathrm{B(int)}}}  \,]\,\,\frac{\cos\theta_{_{\mathrm{B(int)}}}}{n\,\cos\psi_{_{\mathrm {B(int)}}}}\,\,\frac{\delta}{k\,{\mathrm{w}}_{\0}}\\
& =& \frac{2\,\cos\theta_{_{\mathrm{B(int)}}}}{\cos\psi_{_{\mathrm{B(int)}}}}\,\,\frac{\delta}{k\,{\mathrm{w}}_{\0}}\,\,,
 \end{eqnarray*}
 we find the same result obtained for the reflected beam, i.e.
\begin{equation*}
D_{_{\mathrm{UTRA}}}^{^{{\mathrm{[TM]}}}}\left[\theta_{_{\mathrm{B(int)}}}\,+\,\,\displaystyle{\frac{\delta}{k\,{\mathrm{w}}_{\0}}}\right]
\,= \,\frac{k\,{\mathrm{w}}_{\0}}{\delta}
\end{equation*}
and
\begin{equation}
\alpha_{_{\mathrm{UTRA}}}^{^{\mathrm{[TM]}}} \left[\theta_{_{\mathrm{B(int)}}} \,+\,\,\displaystyle{\frac{\delta}{k\,{\mathrm{w}}_{\0}}}\right] \,=\,\frac{2\,\delta}{1+ \delta^{^{2}}}\,\frac{1}{k\,{\mathrm{w}}_{\0}}\,\,.
\end{equation}
Interesting is the curve given  in Fig.\,4(a), where  for a BK7 prism ($n=1.515$) and an incident Gaussian beam with  $\lambda=0.633\,\mu{\mathrm{m}}$ and ${\mathrm{w}}_{\0}=1\,{\mathrm{mm}}$, we show the angular deviations for TM waves in the case of incidence in the internal Brewster region. For example, for incidence at  $\theta_{_{\mathrm{B(int)}}}\,\pm\,1\,/\,k\,{\mathrm{w}}_{\0}$  (incidence for which we find the maximal angular deviation for TM waves) , we have
\[
\begin{array}{l}
\alpha_{_{\mathrm{UTRA}}}^{^{\mathrm{[TE]}}}\,\approx\,\frac{4}{(k\,{\mathrm{w}}_{\0})^{^{2}}}\,\approx \,2.3^{\circ}\,\times\,
10^{^{-6}}\,\,,\\ \\
\alpha_{_{\mathrm{UTRA}}}^{^{\mathrm{[TM]}}}\,\approx\,\pm\, \frac{1}{k\,{\mathrm{w}}_{\0}}\,\approx\,\pm\, \, 5.8^{\circ}\,\times\,
10^{^{-3}}\,\,.
 \end{array}
 \]
{The relative factor of} $k\,{\mathrm{w}}_{\0}/4\approx 2.5\,\times\,10^{^{3}}$ leads to the giant GH angular shift. Before discussing what happens in the critical region, let us call the reader's attention  to the curves plotted in Fig.\,5(a-e) where  the transmitted angular distribution is depicted for different incidence angles. Approaching  the Brewster angle, the angular  profile is strongly distorted by the presence of a secondary peak. Consequently,  the concept of angular shift is obscured \cite{ANG4}. For incidence at $\theta_{_{\mathrm{B(int)}}}\,\pm\,\lambda/{\mathrm{w}}_{\0}$  ($\delta=\pm\, 2\,\pi$), we are still  in the presence of  a  single peak, for the negative $\delta$ case, see Fig.\,4(a), and the concept  of angular deviations can be correctly used. For this incidence angle, we find the reduction factor of  $4\,\pi/[1+(2\,\pi)^{^{2}}]$, approximately $0.31$, still preserving the giant Goos-H\"anchen value. In Fig.\,3(a), the region around the Brewster angle,
 \[  \theta_{_{\mathrm{B(int)}}} -  \,\displaystyle{\frac{\lambda}{\,{\mathrm{w}}_{\0}}}
     < \,\, \theta_{\0} \,<\,\,
    \theta_{_{\mathrm{B(int)}}} + \,\displaystyle{\frac{\lambda}{\,{\mathrm{w}}_{\0}}}\,\,,  \]
where  the interpretation of angular deviation is obscured by the presence of an  additional peak,  is indicated by  using a different background color.

For incidence in the critical region,
\[
\theta_{\0}= \theta_{_{\mathrm{cri}}} \,-\,\, \frac{|\delta|}{k\,{\mathrm{w}}_{\0}}\,\,\,\,\,\,\,\Rightarrow\,\,\,\,\,\,\,\,
\varphi_{\0}= \varphi_{_{\mathrm{cri}}} \,-\,\, \frac{\cos\theta_{_{\mathrm{cri}}}}{n\,\cos\psi_{_{\mathrm{cri}}}}\,\,\frac{|\delta|}{k\,{\mathrm{w}}_{\0}}\,\,,
 \]
we find
\begin{equation*}
\begin{array}{l}
\displaystyle{
\left\{\,
D_{_{\mathrm{UTRA}}}^{^{{\mathrm{[TE]}}}}\left(\theta_{_{\mathrm{cri}}}\,-\,\,\displaystyle{\frac{|\delta|}{k\,{\mathrm{w}}_{\0}}}\right)
\,,\,
D_{_{\mathrm{UTRA}}}^{^{{\mathrm{[TM]}}}}\left(\theta_{_{\mathrm{cri}}}\,-\,\,\displaystyle{\frac{|\delta|}{k\,{\mathrm{w}}_{\0}}}\right)
\,\right\} \,=}\\ \\
 \displaystyle{\sqrt{2}\,\left[\,\frac{2-n^{\2}+2\,\sqrt{n^{\2}-1}}{(n^{\2}-1)\,(n^{\2}+2\,\sqrt{n^{\2}-1})}\,\right]^{^{1/4}}
\left\{\,1\,,\,n^{\2}\,\right\}\,\,\sqrt{\frac{k\,{\mathrm{w}}_{\0}}{|\delta|}}}\,\,.
\end{array}
\end{equation*}
In this case, the amplification of the angular deviations  is proportional to $\sqrt{k\,{\mathrm{w}}_{\0}}$ and depends on the value of the  refractive index,
{
\begin{eqnarray*}
\alpha_{_{\mathrm{UTRA}}}^{^{\mathrm{\{\,TE\,,\,TM\,\}}}} \left(\theta_{_{\mathrm{cri}}} \,-\,\,\displaystyle{\frac{|\delta|}{k\,{\mathrm{w}}_{\0}}}\right)\,\,=\,\,
\frac{f(n)}{\sqrt{|\delta|}}\,\left\{\,1\,,\,n^{\2}\,\right\}\,\,\frac{1}{\,\,\,(k\,{\mathrm{w}}_{\0})^{^{{3/2}}}} \end{eqnarray*}
where
\begin{equation*}
f(n)\,\,=\,\,2\,\sqrt{2}\,\left[\,\frac{2-n^{\2}+2\,\sqrt{n^{\2}-1}}{(n^{\2}-1)\,(n^{\2}+2\,\sqrt{n^{\2}-1})}\,
\right]^{^{1/4}}\,\,.
\end{equation*}
}
 In Fig.\,3(b), for a BK7 prism and an incident Gaussian beam with  $\lambda=0.633\,\mu{\mathrm{m}}$ and ${\mathrm{w}}_{\0}=1\,{\mathrm{mm}}$, we show the angular deviations for TM waves. Same as for the Brewster region, in the critical region for incidence greater that  $\theta_{_{\mathrm{cri}}} -\lambda/{\mathrm{w}}_{\0}$ the concept of angular deviations is also obscured. Indeed,
 the internal reflection coefficient becomes imaginary, see Fig.\,5(f-j), and the interference between the real and complex part has to be considered.  In this case, numerical calculations are required \cite{Break,Axial} and a mixed effect (composite GH shift) is observed\cite{CGH}. For incidence at  $\theta_{_{\mathrm{cri}}} -\lambda/{\mathrm{w}}_{\0}$, the angular distribution is still real, see Fig.4(f), and angular deviations are the only contribution to the optical beam's shift. This incidence angle also gives the maximal angular deviation in the critical region.

A summary of the results presented in this section can be given by calculating, for TE and TM waves, the angular deviations at the borders of region III,
 \[  \theta_{_{\mathrm{B(int)}}} +  \,\displaystyle{\frac{\lambda}{\,{\mathrm{w}}_{\0}}}
     < \,\, \theta_{\0} <
    \theta_{_{\mathrm{cri}}} - \,\displaystyle{\frac{\lambda}{\,{\mathrm{w}}_{\0}}}\,\,,  \]
which is the region where the concept of angular deviations is not obscured by additional peaks or complex angular distributions and still contains a noticeable increase of the angular deviations. For a BK7 prism and an incident Gaussian beam with  $\lambda=0.633\,\mu{\mathrm{m}}$ and ${\mathrm{w}}_{\0}=1\,{\mathrm{mm}}$, we find,
 for incidence at $\theta_{_{\mathrm{B(int)}}} + \, \lambda\,/\, {\mathrm{w}}_{\0}$,
\begin{eqnarray*}
\left\{\,\alpha_{_{\mathrm{UTRA}}}^{^{\mathrm{[TE]}}}\,,\,\alpha_{_{\mathrm{UTRA}}}^{^{\mathrm{[TM]}}}\,\right\} & \approx &   \, \left\{\,\frac{4}{(k\,{\mathrm{w}}_{\0})^{^{2}}}\,,\,\frac{1/\pi}{k\,{\mathrm{w}}_{\0}}\,\right\} \\
& \approx & \left\{\,2.3^{\circ}\times\,10^{^{-3}}\,,\,1.8^{\circ}\,\right\}\,\times\,10^{^{-3}}\,\,,
\end{eqnarray*}

and, using $f(1.515)\approx 3/\sqrt{2}$, for incidence  at $\theta_{_{\mathrm{cri}}} - \, \lambda\,/\, {\mathrm{w}}_{\0}$,
\begin{eqnarray}
\left\{\,\alpha_{_{\mathrm{UTRA}}}^{^{\mathrm{[TE]}}}\,,\,\alpha_{_{\mathrm{UTRA}}}^{^{\mathrm{[TM]}}}\,\right\} & \approx &
\frac{3}{2\,\sqrt{\pi}}\,\{\,1\,,\,2.3\,\}\, \frac{1}{\,\,(k\,{\mathrm{w}}_{\0})^{^{3/2}}} \\
& \approx &\left\{\,0.5^{\circ}\,,\,1.1^{\circ}\,\right\}\,\times\,10^{^{-4}}\,\,.
\end{eqnarray}
This clearly shows that the local maximum near the critical region is greater than the local maximum near the Brewster region, for both TE and TM waves and. The angular deviations for intermediate angles are given in Table 1.

Before concluding this section, we observe that the angular deviations of the transmitted beam are generated by the angular deviations of the beam reflected at the lower (dielectric/air) interface. To link these angular deviations, let us consider the refracted beam at the left (air/dielectrice) interface. It forms an angle $\varphi_{_{\0}}=\psi_{\0}+\pi/4$ with the normal to the lower interface, see Fig.\,1. The beam reflected at the this interface, deflected by the angular deviations discussed in this section, forms an angle $\widetilde{\varphi_{\0}}=\varphi_{_{\0}}+\delta\varphi$ with the normal to the lower interface. The reflected beam, reaching the right (dielectric/air) interface, forms and angle $\widetilde{\psi_{\0}}=\psi_{_{\0}}+\delta\varphi$ with the normal to this interface. Finally, the angle, $\widetilde{\theta_{\0}}= \theta_{_{\0}}+\delta\theta$, that the transmitted beam forms with the normal at the right interface is calculated by the Snell law, $\sin\widetilde{\theta_{\0}}=n\,\sin\widetilde{\psi_{\0}}$. After simple algebraic manipulations, we finally find that the angular deviations induced on the transmitted beam by the reflected one are given by $\delta\theta=n\,\cos\psi_{\0}\,\delta\varphi/\,\cos\theta_{\0}$.

\subsection{ Lower transmitted beam}

{Following the same procedure as done for the reflected beam we calculate the angular deviation of the lower transmitted beam as
\begin{equation}
\label{meantra}
\langle\, y_{_{\mathrm{LTRA}}}^{^{\mathrm{[TE,TM]}}} \rangle
  = y_{_{\mathrm{LGEO}}} +\alpha_{_{\mathrm{LTRA}}}^{^{\mathrm{[TE,TM]}}}(\theta_{\0})\,\,\,(z_{_{\mathrm{LTRA}}}-z_{_{\mathrm{LGEO}}})
\end{equation}
where
\begin{equation}
\alpha_{_{\mathrm{LTRA}}}^{^{\mathrm{[TE,TM]}}}(\theta_{\0})\,=\,
\phi_{\0}^{\prime}\,\frac{2\, D_{_{\mathrm{LTRA}}}^{^{{\mathrm{[TE,TM]}}}}(\theta_{\0})}{(k\,{\mathrm{w}}_{\0})^{^{^2}}\,+\, [\,D_{_{\mathrm{LTRA}}}^{^{{\mathrm{[TE,TM]}}}}(\theta_{\0})\,]^{^{^{^2}}}}
\end{equation}
and
\begin{eqnarray}
D_{_{\mathrm{LTRA}}}^{^{{\mathrm{[TE,TM]}}}}(\theta_{\0})  &=& T_{_{\mathrm{down}}}^{^{{\mathrm{[TE,TM]}}^{\prime}}}(\theta_{\0})\bigg/T_{_{\mathrm{down}}}^{^{{\mathrm{[TE,TM]}}}}(\theta_{\0}) {\nonumber}\\
&=&\begin{array}{l}
{\displaystyle \frac{n^{^2}-1}{n}\,\frac{\tan\varphi_{\0}\,\cos\theta_{\0}}{\cos\phi_{\0}\,\cos\psi_{\0}}\,\,\times}\\
\displaystyle{\left\{\, \frac{1}{n\,\cos\varphi{\0}+\cos\phi_{\0}} \,,\, \frac{n}{\cos\varphi{\0}+n\,\cos\phi_{\0}}  \,\right\}}\,\,.
\end{array}
\end{eqnarray}
Note that $T_{_{\mathrm{left}}}^{^{^{\mathrm{[TE,TM]}}}}$ is a smooth function and consequently the main contribution to angular deviations comes from $T_{_{\mathrm{down}}}^{^{^{\mathrm{[TE,TM]}}}}$.

For incidence in the critical region,
we find
\begin{equation*}
\begin{array}{l}
\displaystyle{
\left\{\,
D_{_{\mathrm{LTRA}}}^{^{{\mathrm{[TE]}}}}\left(\theta_{_{\mathrm{cri}}}\,-\,\,\displaystyle{\frac{|\delta|}{k\,{\mathrm{w}}_{\0}}}\right)
\,,\,
D_{_{\mathrm{LTRA}}}^{^{{\mathrm{[TM]}}}}\left(\theta_{_{\mathrm{cri}}}\,-\,\,\displaystyle{\frac{|\delta|}{k\,{\mathrm{w}}_{\0}}}\right)
\,\right\} \,=}\\ \\
 \displaystyle{\frac{1}{\sqrt{2}}\,\left[\,\frac{2-n^{\2}+2\,\sqrt{n^{\2}-1}}{(n^{\2}-1)\,(n^{\2}+2\,\sqrt{n^{\2}-1})}\,\right]^{^{1/4}}
\left\{\,1\,,\,n^{\2}\,\right\}\,\,\sqrt{\frac{k\,{\mathrm{w}}_{\0}}{|\delta|}}}\,\,
\end{array}
\end{equation*}
and
\begin{equation*}
\begin{array}{l}
 \phi_{\0}^{\prime}\left(\theta_{_{\mathrm{cri}}}\,-\,\,\displaystyle{\frac{|\delta|}{k\,{\mathrm{w}}_{\0}}}\right)\,=
 \displaystyle{\frac{1}{\sqrt{2}}\,\left[\,\frac{(n^{\2}-1)\,(2-n^{\2}+2\,\sqrt{n^{\2}-1})}{n^{\2}+2\,\sqrt{n^{\2}-1}}\,\right]^{^{1/4}}
  \,\,\sqrt{\frac{k\,{\mathrm{w}}_{\0}}{|\delta|}}}\,\,.
\end{array}
\end{equation*}

In the vicinity of the critical angle, considering an incident angle of $\theta_{\0} = \theta_{_{\mathrm{cri}}}-|\delta|/k\mathrm{w}_{\0}$, the angular deviations are given by
\begin{eqnarray*}
\alpha_{_{\mathrm{LTRA}}}^{^{\mathrm{\{\,TE\,,\,TM\,\}}}} \left(\theta_{_{\mathrm{cri}}} \,-\,\,\displaystyle{\frac{|\delta|}{k\,{\mathrm{w}}_{\0}}}\right)\,\,=\,\,
 \frac{\sqrt{n^{\2}-1}\,f^{\2}(n)}{8}\,\left\{\,1\,,\,n^{\2}\,\right\}\,\,\frac{1}{\,\,\,|\delta|\,k\,{\mathrm{w}}_{\0}}\,\,.\end{eqnarray*}
At the threshold of the critical region, that is, for $\delta=2\pi$, we have the maximal angular deviation before an appreciable part of the beam enters the total internal reflection regime.
By using $\sqrt{1.515^{\2}-1}\,f^{\2}(1.515)\approx \sqrt{26}$, for incidence  at $\theta_{_{\mathrm{cri}}} - \, \lambda\,/\, {\mathrm{w}}_{\0}$,
\begin{eqnarray}
\left\{\,\alpha_{_{\mathrm{LTRA}}}^{^{\mathrm{[TE]}}}\,,\,\alpha_{_{\mathrm{LTRA}}}^{^{\mathrm{[TM]}}}\,\right\} &\approx &
\frac{\sqrt{26}}{16\,\pi}\,\{\,1\,,\,2.3\,\}\, \frac{1}{\,\,k\,{\mathrm{w}}_{\0}} \\
&\approx &
\left\{\,0.6^{\circ}\,,\,1.3^{\circ}\,\right\}\,\times\,10^{^{-3}}\,\,.
\end{eqnarray}
 }

\section{Amplification by optical weak measurements}

Up to this point, by presenting  a direct measurement of angular deviations, we have limited ourselves to the strict classic approach of the problem. The results presented in the previous section show that,  for incidence close to the critical angle, a new amplification region is present. This  amplification is  of the order of $\sqrt{k\,{\mathrm{w}}_{\0}}$. Nevertheless, it is  lower than the standard Brewster amplification which is of the order of $k\,{\mathrm{w}}_{\0}$.   For a  measuring procedure based on the weak measurement technique, this situation drastically changes leading to what  we call the breaking off of the giant Goos-H\"anchen angular shift. For the critical region, the weak measurement  power amplification still works  and for appropriate choices of the weak measurement parameters  it is possible to optimize such an amplification reverting the results of the direct measurement.

Let us now consider  a weak measurement set-up as the one represented in Fig.\,6(a). We start with a brief description of this measuring procedure.  Before reaching the dielectric prism the laser passes through a first polarizer with angle $\pi/4$. This creates   an equal mixture of $\mathrm{TE}$  and $\mathrm{TM}$  waves. Then, after interacting with the dielectric, the outgoing beam passes through a second polarizer with an angle $\beta$ and  is finally  detected by the  camera. The intensity measured at the camera is thus given by
\begin{equation}
I_{_{\mathrm{UTRA}}} \,\propto\,\left|\,\sin\beta \, E_{_{\mathrm{UTRA}}}^{^{\mathrm{[TE]}}}  \, + \,\, \cos\beta \, E_{_{\mathrm{UTRA}}}^{^{\mathrm{[TM]}}}\,\right|^{^2}\,\,,
\end{equation}
where
\begin{equation*}
E_{_{\mathrm{UTRA}}}^{^{\mathrm{[TE,TM]}}}\,
\propto \,\,\,  R^{^{\mathrm{[TE,TM]}}}_{_{\mathrm{down}}}(\theta_{\0}) \,\,
\mathrm{exp}\left\{\, -\,\frac{\displaystyle \left[\,\widetilde{y}_{_{\mathrm{UTRA}}} -\,\, \alpha_{_{\mathrm{UTRA}}}^{^{\mathrm{[TE,TM]}}}(\theta_{\0}) \,\,
\widetilde{z}_{_{\mathrm{UTRA}}}\right]^{^2}}{\displaystyle {\mathrm{w}}^{\2}(\widetilde{z}_{_{\mathrm{UTRA}}})} \right\}
\end{equation*}
and $ {\mathrm{w}}(\widetilde{z}_{_{\mathrm{UTRA}}})  \, = \, {\mathrm{w}}_{\0}\, \sqrt{1 + \left( \lambda \, \widetilde{z}_{_{\mathrm{UTRA}}}  \,/\,\pi \, {\mathrm{w}}_{\0}^{^2} \right)^{^2}}$.  By introducing the dimensionless quantities
\begin{eqnarray*}
Y \, = \, \left[\,\widetilde{y}_{_{\mathrm{UTRA}}}- \,\frac{  \alpha_{_{\mathrm{UTRA}}}^{^{\mathrm{[TE]}}}(\theta_{\0}) + \, \alpha_{_{\mathrm{UTRA}}}^{^{\mathrm{[TM]}}}(\theta_{\0}) }{2} \,\,\, \widetilde{z}_{_{\mathrm{UTRA}}}\,\right]\,/\,{\mathrm{w}}(\widetilde{z}_{_{\mathrm{UTRA}}})   \,\,\,,\\
 Z=  \widetilde{z}_{_{\mathrm{UTRA}}}\,/\,{\mathrm{w}}(\widetilde{z}_{_{\mathrm{UTRA}}}) \,\,,
\end{eqnarray*}
and
\begin{eqnarray*}
\Delta\alpha_{_{\mathrm{UTRA}}}(\theta_{\0}) \, = \,\alpha_{_{\mathrm{UTRA}}}^{^{\mathrm{[TM]}}}(\theta_{\0}) - \, \alpha_{_{\mathrm{UTRA}}}^{^{\mathrm {[TE]}}} (\theta_{\0})\,\,\,\\ \\
\tau(\theta_{\0}) = R^{^{\mathrm{[TM]}}}_{_{\mathrm{down}}}(\theta_{\0})\,/\,R^{^{\mathrm{[TE]}}}_{_{\mathrm{down}}}(\theta_{\0}) \,\,,
\end{eqnarray*}
we can rewrite  the intensity as follows
\begin{eqnarray*}
I_{_{\mathrm{UTRA}}} \, \propto \, \left\{ \,\tan\beta \, \exp\left[\,-\,\left( Y + \,\frac{\Delta\alpha_{_{\mathrm{UTRA}}}(\theta_{\0})}{2} \,\, \,Z\, \right)^{^2}\, \right] +
\tau(\theta_{\0})\, \exp\left[\,-\,\left( Y - \,\frac{\Delta\alpha_{_{\mathrm{UTRA}}}(\theta_{\0})}{2} \,\, \,Z\, \right)^{^2} \,\right]
 \,\right\}^{^2}\,\,.
\end{eqnarray*}
Let us now set  the second polarizer to $\beta  \, = \, -\,\arctan[\,\tau(\theta_{\0})\,] \,+\,\Delta\epsilon$, with  $|\Delta \epsilon| \ll 1$. This choice implies
\[
\tan \beta \,\approx \,-\,\tau(\theta_{\0})\,+\,\left[\,1 \,+\,\tau^{^{2}}(\theta_{\0})  \,\right]\,\Delta \epsilon\,\,.
\]
Using the previous result and observing that  $\Delta\alpha_{_{\mathrm{UTRA}}}(\theta_{\0}) \ll 1$, we can further  simplify the transmitted  intensity,
\begin{eqnarray}
\label{IntWM}
I_{_{\mathrm{UTRA}}} & \propto & \left\{ \,[\,-\,\tau(\theta_{\0})+(1+\tau^{\2}(\theta_{\0}))\,\Delta \epsilon\,] \, (\,1- \Delta\alpha(\theta_{\0})\,Y\,Z\,) + \tau(\theta_{\0})\,(\,1+ \Delta\alpha(\theta_{\0})\,Y\,Z\,)\,\right\}^{^{2}}\,\exp[\,-\,2\,Y^{^{2}}\,]\nonumber \\
 & \propto &  \left[ \,\frac{1+\tau^{\2}(\theta_{\0})}{2\, \tau(\theta_{\0})}\,\Delta \epsilon\, +\,\Delta\alpha(\theta_{\0})\,Y\,Z\,\right]^{^{2}} \,\exp[\,-\,2\,Y^{^{2}}\,]\nonumber \\
 & = &  \left[ \,\frac{\Delta\epsilon}{A(\theta_{\0})}\, +\,\Delta\alpha(\theta_{\0})\,Y\,Z\,\right]^{^{2}} \,\exp[\,-\,2\,Y^{^{2}}\,]\,\,.
\end{eqnarray}
This  transmitted beam profile is characterized by two peaks located at
\begin{equation}
Y_{_{\mathrm{max}}}^{^{\pm}}(\Delta \epsilon) = \frac{-\,\Delta\epsilon\,\pm\,\sqrt{(\Delta\epsilon)^{^{2}} + 2\,[\,A(\theta_{\0})\,\Delta\alpha(\theta_{\0})\,Z\,\,]^{^{2}}}}{2\,A(\theta_{\0})\,\Delta\alpha(\theta_{\0})\,Z}\,\,.
\end{equation}
If $|\Delta\epsilon|\,\gg\,A(\theta_{\0})\,\Delta\alpha(\theta_{\0})\,Z$ (we come back to the implications of this constraint later), we can approximate  the square root by
\[|\Delta\epsilon| \,\,+\,\,\frac{[\,A(\theta_{\0})\,\Delta\alpha(\theta_{\0})\,Z\,\,]^{^{2}}}{|\Delta \epsilon|}\,\,. \]
For a positive rotation of the second polarizer, $\Delta \epsilon = |\Delta \epsilon |$, using the previous approximation we  find
\begin{eqnarray}
\left\{\,Y_{_{\mathrm{max}}}^{^{-}}(|\Delta \epsilon|)\,,\, Y_{_{\mathrm{max}}}^{^{+}}(|\Delta \epsilon|)\,\right\}\,=\,
\left\{\,-\,\frac{|\Delta \epsilon|}{A(\theta_{\0})\,\Delta\alpha(\theta_{\0})\,Z}\,,\,  \frac{A(\theta_{\0})\,\Delta\alpha(\theta_{\0})\,Z}{2\,|\Delta \epsilon|}  \,\right\}\,\,.
\end{eqnarray}
This shows that for a  positive rotation the main peak is found at $Y_{_{\mathrm{max}}}^{^{+}}(|\Delta \epsilon|)$. By repeating the measurement for a negative rotation, $\Delta \epsilon = -\,|\Delta \epsilon |$, we find
\begin{eqnarray}
\left\{\,Y_{_{\mathrm{max}}}^{^{-}}(-\,|\Delta \epsilon|)\,,\, Y_{_{\mathrm{max}}}^{^{+}}(-\,|\Delta \epsilon|)\,\right\}\,=\,
\left\{\,  -\, \frac{A(\theta_{\0})\,\Delta\alpha(\theta_{\0})\,Z}{2\,|\Delta \epsilon|}\,,\, \frac{|\Delta \epsilon|}{A(\theta_{\0})\,\Delta\alpha(\theta_{\0})\,Z}\,,\,  \,\right\}\,\,.
\end{eqnarray}
In this case, the main peak is found at $Y_{_{\mathrm{max}}}^{^{-}}(-\,|\Delta \epsilon|)$. The difference between these peaks,
\begin{eqnarray}
\Delta Y_{_{\mathrm{max}}} \,=\,  Y_{_{\mathrm{max}}}^{^{+}}(|\Delta \epsilon|) - Y_{_{\mathrm{max}}}^{^{-}}(-\,|\Delta \epsilon|) \,=\, \frac{ A(\theta_{\0})}{|\Delta \epsilon\,|}\,\,\Delta\alpha_{_{\mathrm{UTRA}}}(\theta_{\0})\,\,Z = \Delta\alpha_{_{\mathrm{UTRA}}}^{^{\mathrm{WM}}}(\theta_{\0})\,\,Z\,\,,
\end{eqnarray}
is what is detected in a weak measurement experiment. With respect to a direct measuring procedure, $\Delta\alpha_{_{\mathrm{UTRA}}}(\theta_{\0})$, the weak measurement angular deviations, $\Delta\alpha_{_{\mathrm{UTRA}}}^{^{\mathrm{WM}}}(\theta_{\0})$,
 contain the amplification factor $1\,/\,|\Delta \epsilon|$.  For incidence approaching the critical angle, $A_{_{\mathrm{cri}}}(\theta_{\0})\approx 1$,  this amplification represents the effective amplification of the angular deviations. So, in the critical region, we have
\begin{equation}
\Delta\alpha_{_{\mathrm{UTRA, cri}}}^{^{\mathrm{WM}}}(\theta_{\0})\,\, \approx\,\,
\frac{\Delta\alpha_{_{\mathrm{UTRA, cri}}}(\theta_{\0})}{|\Delta \epsilon|}\,\,\propto\,\, \frac{1}{|\Delta \epsilon|\,\,(k\,{\mathrm{w}}_{\0})^{^{3/2}}}\,\,.
\end{equation}
In the Brewster region, the factor $A_{_{\mathrm{B(int)}}}(\theta_{\0})$ is proportional to $1/\,k\,{\mathrm{w}}_{\0}$ and, consequently, creates
an anti-giant effect compensating the amplification of the  direct measurement,
\begin{equation}
\Delta\alpha_{_{\mathrm{UTRA, B(int)}}}^{^{\mathrm{WM}}}(\theta_{\0})\,\, \propto\,\,
\frac{\Delta\alpha_{_{\mathrm{UTRA, B(int)}}}(\theta_{\0})}{k\,{\mathrm{w}}_{\0}\,|\Delta \epsilon|}\,\,\propto\,\, \frac{1}{|\Delta \epsilon|\,\,(k\,{\mathrm{w}}_{\0})^{^{2}}}\,\,.
\end{equation}
The rotation for which a weak measurement changes the amplification power in the Brewster and critical regions  is given by  $|\Delta \epsilon|=1/\,\sqrt{k\,{\mathrm{w}}_{\0}}$. For a laser with wavelength $\lambda=0.633\,\mu{\mathrm{m}}$ and ${\mathrm{w}}_{\0}=1\,{\mathrm{mm}}$, such a rotation  corresponds to an angle  of $0.575^{^{0}}$.

Let us now come back to the constraint
\begin{equation}
|\Delta\epsilon|\,\gg\,A(\theta_{\0})\,\Delta\alpha(\theta_{\0})\,Z
\end{equation}
and calculate the minimal value of the second polarizer rotation which can be used in the weak measurement analysis.
To set a common value of the rotation angle in region III,
$\theta_{_{\mathrm{B(int)}}}+\lambda/{\mathrm{w}}_{\0}<\theta_{\0}< \theta_{_{\mathrm{cri}}}-\lambda/{\mathrm{w}}_{\0}$,
we observe that the main restriction comes from the critical region where $A(\theta_{\0})\approx 1$ and the angular deviations are proportional to $(k\,{\mathrm{w}}_{\0})^{^{-\,3/2}}$. Consequently, we have
\begin{equation}
|\Delta\epsilon|\,\gg\,  \frac{Z}{\,\,\,(k\,{\mathrm{w}}_{\0})^{^{3/2}}}\,\,.
\end{equation}
For a beam with ${\mathrm{w}}_{\0}=1\,{\mathrm{mm}}$, $\lambda=0.633\,\mu{\mathrm{m}}$ and a camera positioned at $z_{_{\mathrm{UTRA}}}=25\,cm$, we have
\[z_{_{\mathrm{UGEO}}}\,\ll\, z_{_{\mathrm{UTRA}}}\,\ll\, k\,{\mathrm{w}}_{\0}^{^{2}}\,\,\,\,\,\,\,\,\,\Rightarrow\,\,\,\,\,\,\,\,\,\,Z\,\approx\,z_{_{\mathrm{UTRA}}}/\,{\mathrm{w}}_{\0}\,=\,250\,\,. \]
Observing that  $k\,{\mathrm{w}}_{\0}\approx 10^{^{4}}$, we obtain the following constraint
\[ |\Delta \epsilon|\,\gg\,2.5\,\times\,10^{^{-\,4}}\,\,(\,\approx 0.014^{\circ}\,)\,\,.\]
In Fig.\,6(b), we show the amplification for a weak measurement of angular deviations for different rotation angles of the second polarizer, $|\Delta \epsilon|=0.1^{\circ},\,\,0.2^{\circ}$ and $0.5^{\circ}$.
In the insets (c) and (d), we respectively zoom to the internal Brewster and critical region. In (c) the breaking off of  weak measurements is clear. Approaching the critical region, inset (d), the WM amplification is evident. At the borders of region III, the comparison between a direct measurement of angular deviations and its weak measurement  counterpart for $|\Delta \epsilon|=0.1^{\circ}$,
\begin{eqnarray*}
\left\{\,\Delta\alpha_{_{\mathrm{UTRA}}}\,,\,\Delta\alpha_{_{\mathrm{UTRA}}}^{^{\mathrm{Weak}}}  \,\right\} &=& \left\{\,1.8^{\circ} \,,\,\,\,\,3.0^{\circ}\,\right\}\,\times\,10^{^{-3}}
\end{eqnarray*}
at $\theta_{_{\mathrm{B(int)}}} + \, \lambda\,/\, {\mathrm{w}}_{\0}$ and
\begin{eqnarray*}
\left\{\,\Delta\alpha_{_{\mathrm{UTRA}}}\,,\,\Delta\alpha_{_{\mathrm{UTRA}}}^{^{\mathrm{Weak}}}  \,\right\} &=& \left\{\,6.5^{\circ}\,\times\,10^{^{-2}} \,,\,37.0^{\circ}\,\right\}\,\times\,10^{^{-3}}
\end{eqnarray*}
at $\theta_{_{\mathrm{cri}}} - \, \lambda\,/\, {\mathrm{w}}_{\0}$
clearly shows the amplification power of the weak measurement technique. For intermediate angles, the comparison between direct and weak measuring procedures  is given in Table 1.

\section{Conclusions}

{ The beam shifts field has been a matter of great interest and widely studied. Still, its richness allow for new developments and new perspectives on earlier research. In the present work we have revisited the topic of angular deviations for a Gaussian beam interacting with a dielectric triangular prism with a focus on the weak measurements and an analysis of its efficiency in different regions of incidence angles.} For incidence close to the internal Brewster angle we reproduce the same results of the external reflection but, in addition to the { deviation peak} of the Brewster region, we find a {\em new}  region { of large angular deviation phenomenon} for incidence close the critical angle. The analytic description given in section III shows that  in the  Brewster region the amplification of the angular deviations is of the order of $k\,{\mathrm{w}}_{\0}$, leading to the giant Goss-H\"anchen angular shift, and in the critical region  the  amplification is of the order of  $\sqrt{k\,{\mathrm{w}}_{\0}}$ what we can nickname   a $\sqrt{\mathrm{giant}}$\,\,Goos-H\"anchen effect.
In the critical region,  polarization also acts as a scale factor. A $\mathrm{TM}$-polarized beam is more effectively deviated (factor  $n^{{\2}}$)  than a $\mathrm{TE}$-polarized beam. The structure of the beam is also relevant  because the wider its angular distribution is the more susceptible it will be to the drastic angular variation imposed by the transmission coefficient and, consequently, the more intense will be its deviation from the geometrical predictions. As we narrow the angular distribution, forcing the beam into the plane wave limit, the angular deviations start to fade. Angular deviations are neither found as we move to the right of the critical angle because, in the total internal reflection regime, the transmission coefficient becomes constant in modulus, thus  preserving the beam's symmetry. Since an incidence at the critical angle would produce the most asymmetric beam, it would be expected to be the incidence of maximal deviation. However, for incidence at the critical angle the angular distribution, the part with $\theta>\theta_{_{\mathrm{cri}}}$ becomes complex  and a new effect appears mixing the angular deviations with the Goos-H\"anchen lateral shift, generating the so-called composite Goos-H\"anchen effect\cite{CGH}. The analytical results presented in this paper refer to the case of a real angular distribution, this means incidence in the region
\[  \theta_{_{\mathrm{B(int)}}} +  \,\displaystyle{\frac{\lambda}{\,{\mathrm{w}}_{\0}}}
     < \,\, \theta_{\0} <
    \theta_{_{\mathrm{cri}}} - \,\displaystyle{\frac{\lambda}{\,{\mathrm{w}}_{\0}}}\,\,.  \]
For a direct measurement of angular deviations the Brewster region is preferred with respect to the critical one. Observing that,  the shift caused on the beam by the angular deviations in the Brewster region is proportional to $z/k\,{\mathrm{w}}_{\0}$ while the one in the critical region is proportional to
$z/(k\,{\mathrm{w}}_{\0})^{^{3/2}}$, to quantify the efficiency of a direct measurement, we can introduce as a   quantifier the ratio  between two beam parameters,  the shift of the peak at $z$  and its width ${\mathrm{w}}(z)$,

\begin{eqnarray}
\rho_{_{\mathrm{B(int)}}}& = & \frac{z}{k\,{\mathrm{w}}_{\0}\,{\mathrm{w}}(z)}\,\,,\nonumber \\
\rho_{_{\mathrm{cri}}}&= & \frac{z}{(k\,{\mathrm{w}}_{\0})^{^{3/2}}\,{\mathrm{w}}(z)}\,\,.
\end{eqnarray}

{\noindent}For example, for the beam considered in this paper at $z=25$ cm, we find
\[
\left\{\,\rho_{_{\mathrm{B(int)}}}\,,\,\rho_{_{\mathrm{cri}}}\,\right\}\,=\,\{\,2.515\,\%\,,\,0.025\,\%\,\}\,\,.
\]
For a weak measurement, this efficiency factor is given by the dimensionless quantity $\Delta Y_{_{\mathrm{max}}}$,
\begin{eqnarray}
\rho_{_{\mathrm{B(int),WM}}} & = & \frac{z}{|\Delta \epsilon|\,(\,k\,{\mathrm{w}}_{\0}\,)^{^{2}}\,{\mathrm{w}}(z)}\,\,,\nonumber \\
\rho_{_{\mathrm{cri,WM}}}& = & \frac{z}{|\Delta \epsilon|\,(k\,{\mathrm{w}}_{\0})^{^{3/2}}\,{\mathrm{w}}(z)}\,\,.
\end{eqnarray}
For a second polarizer rotation of $|\Delta \epsilon|= 1/\sqrt{k\,{\mathrm{w}}_{\0}}$, we invert the efficiency obtained by a direct measurement and  for
 \[   \frac{z}{(\,k\,{\mathrm{w}}_{\0})^{^{3/2}}\,{\mathrm{w}}(z)}\,\ll\,  |\Delta \epsilon |\,<\,  \frac{1}{\sqrt{k\,{\mathrm{w}}_{\0}}}\,\,, \]
we improve the angular deviations amplification of a direct measuring procedure.  The constraints on $\Delta \epsilon$ immediately suggest that for a  fixed $z$, the larger the beam width is, the larger the interval of angles becomes.
For the beam considered in this paper and the camera positioned at $z=25\,{\mathrm{cm}}$, the $|\Delta \epsilon|$ constraint  becomes
\[ 2.5\,\times\,10^{^{-4}}\,\ll\,|\Delta \epsilon|\,<\,10^{^{-2}}
\,\,\,\,\,\,\,\,\,
\Rightarrow
\,\,\,\,\,\,\,\,\,
0.014^{\circ}\,\ll\,|\Delta \epsilon|^{\circ}\,<\,0.573^{\circ}\,\,,
 \]
and fixing the rotation of the second polarizer at $|\Delta\epsilon|=0.1^{\circ}$, in the critical region, we gain, with respect to a direct measurement, an amplification factor of $1800/\pi$. In the Brewster region, we have an anti-amplification effect proportional to $0.18/\pi$,
\begin{eqnarray*}
\left\{\,\rho_{_{\mathrm{B(int)}}}\,,\,\rho_{_{\mathrm{cri}}}\,\right\} & : &
\{\,2.515\,\%\,,\,0.025\,\%\,\}_{_{\mathrm{DM}}}\\
 & &
\{\,0.145\,\%\,,\,14.466\,\%\,\}_{_{\mathrm{WM}}}\,\,,
\end{eqnarray*}
which we  call the weak measurement breaking off of the giant GH angular shift.

Finally, { we would like to emphasize reason for the system of our choice. Weak measurements of angular deviations have been mostly performed for external reflections \cite{ExpWM2,ExpWM3} but internal reflections also provide an interesting set-up. It allows us to compare the technique's performance in two different regions where two different symmetry breaking process take place. This is an advantage in relation to external reflections, where only the Brewster region is available, because it tests the relative efficiency of the weak measurement, revealing it is not equally efficient in measurements carried out in the incidence region between the internal Brewster and the critical regions.}

We conclude our paper by observing that the weak measurement surely represents one of the most efficient methods to amplify deviations from the geometrical optics. We hope that the analytical expression given for the angular deviations, the new amplification region found in the proximity of the critical incidence, and  the detailed analysis of the weak measuring procedure  will be useful to stimulate further theoretical studies, as well as to  realize new weak measurement experimental investigations.\\

{{
{\textbf ACKNOWLEDGEMENTS.}
The authors would like to thank the referees and editor for
their suggestions  and their challenging questions. Their observations have been very helpful and of great value to improve the scientific content of the article. The paper in its present form is surely the result of their attentive reading of the manuscript and their stimulating comments.
}}

\newpage

\WideFigure{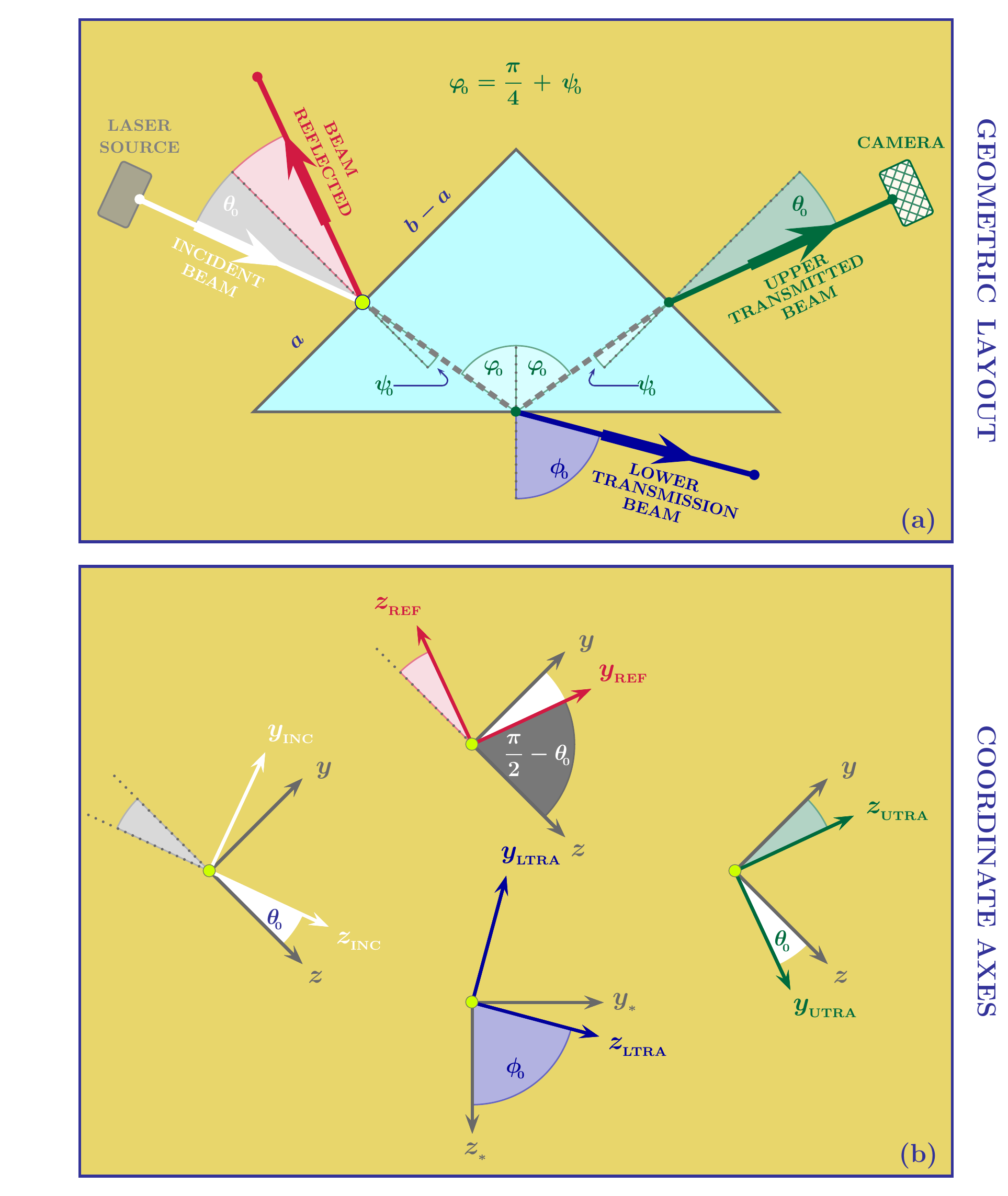}{Planar view of the dielectric block used for the analysis carried out in this paper. The incidence angle at the lower (dielectric/air) interface, $\varphi_{\0}$, is always positive and given by $\pi_{\0}/4+\psi_{\0}$ ($\sin\theta_{\0}=n\,\sin \psi_{\0}$). In (b), the axes describing the propagation  of the incident (INC), reflected (REF) and transmitted (TRA) beams are displayed.}
\newpage

\WideFigure{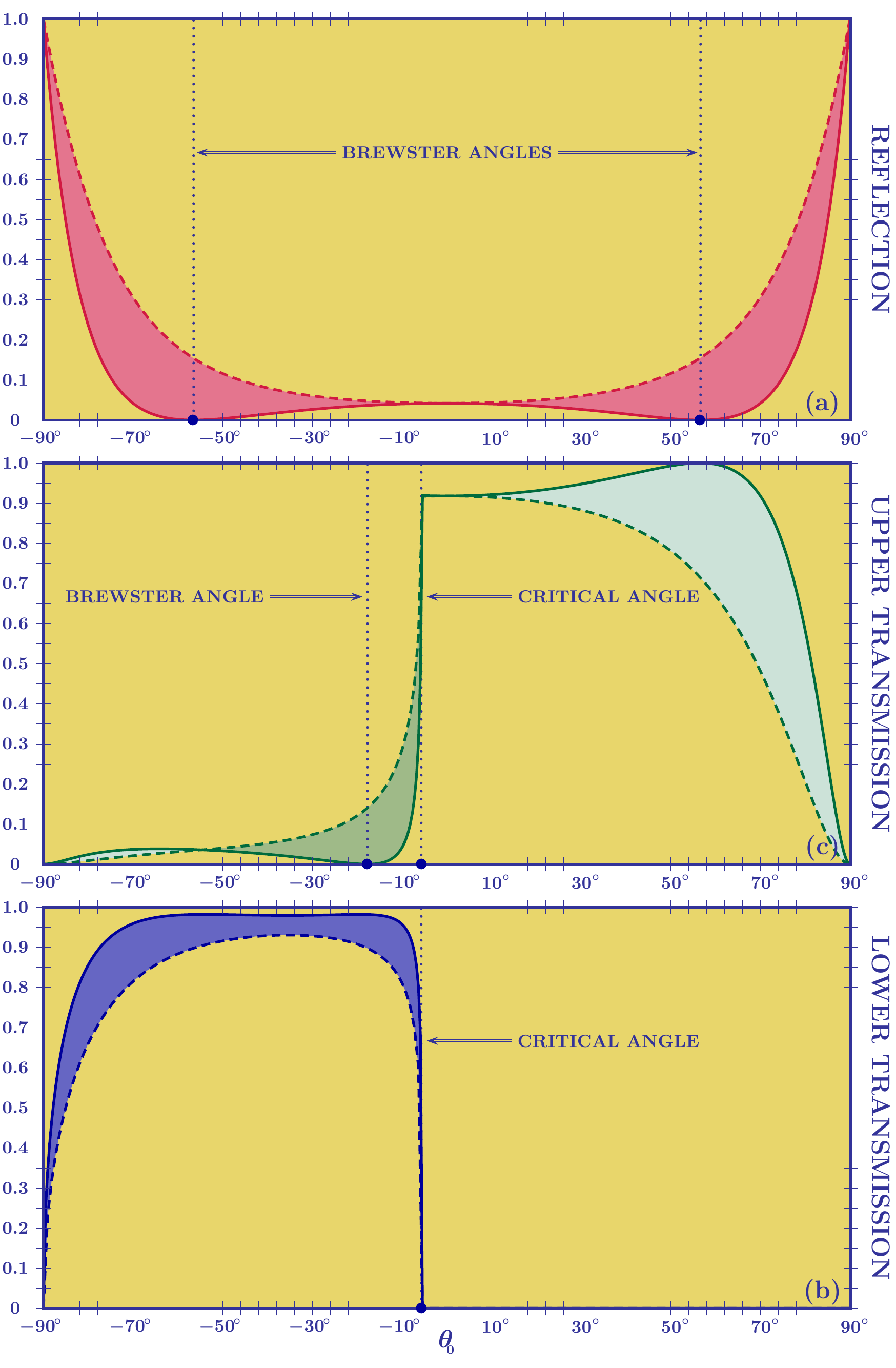}{The reflected (a), upper transmitted (b) and lower transmitted (c) normalized powers as a function of the incidence angle. For external reflection we find two Brewster angles located at $\theta_{\0}=\pm\,56.573^{^{0}}$ while for the internal reflection we only find one Brewster angle before the critical region, $\theta_{\0}=-\,17.693^{\circ}$. The critical angle for internal reflection is found at
$\theta_{\0}=-\,5.603^{\circ}$. The dashed and solid lines represent the Fresnel coefficients for TE and TM waves, respectively. The filled areas between curves emphasize the difference between the Fresnel reflection coefficients for TE and TM waves for a given incident angle.}
\newpage

\WideFigure{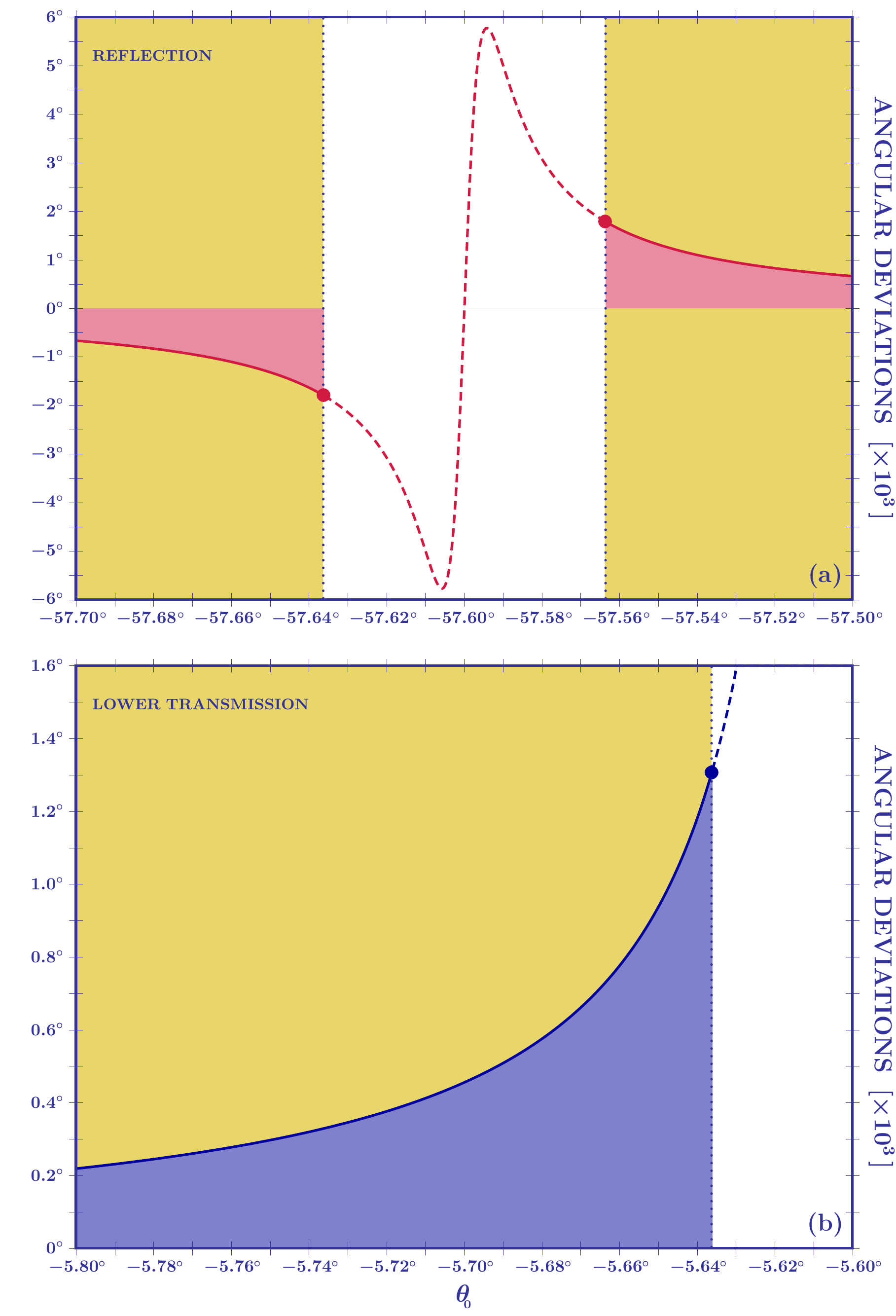}{For a BK7 prism and incident laser beam with $\lambda=0.633\,\mu{\mathrm{m}}$ and waist ${\mathrm{w}}_{\0}=1\,{\mathrm{mm}}$,
the Brewster region for external reflection and the critical region for the lower transmission are plotted in (a) and (b), respectively.  In (a) the amplification is proportional $k\,{\mathrm{w}}_{\0}$ producing the so-called giant GH angular shift while in (b) it is proportional to
 $1/k\,{\mathrm{w}}_{\0}$. The white background indicates the region where the concept of angular deviation is obscured due to the presence of an additional peak in the angular distribution of the reflected beam (a) or to the complex nature of the reflection coefficient (b).}
\newpage

\WideFigure{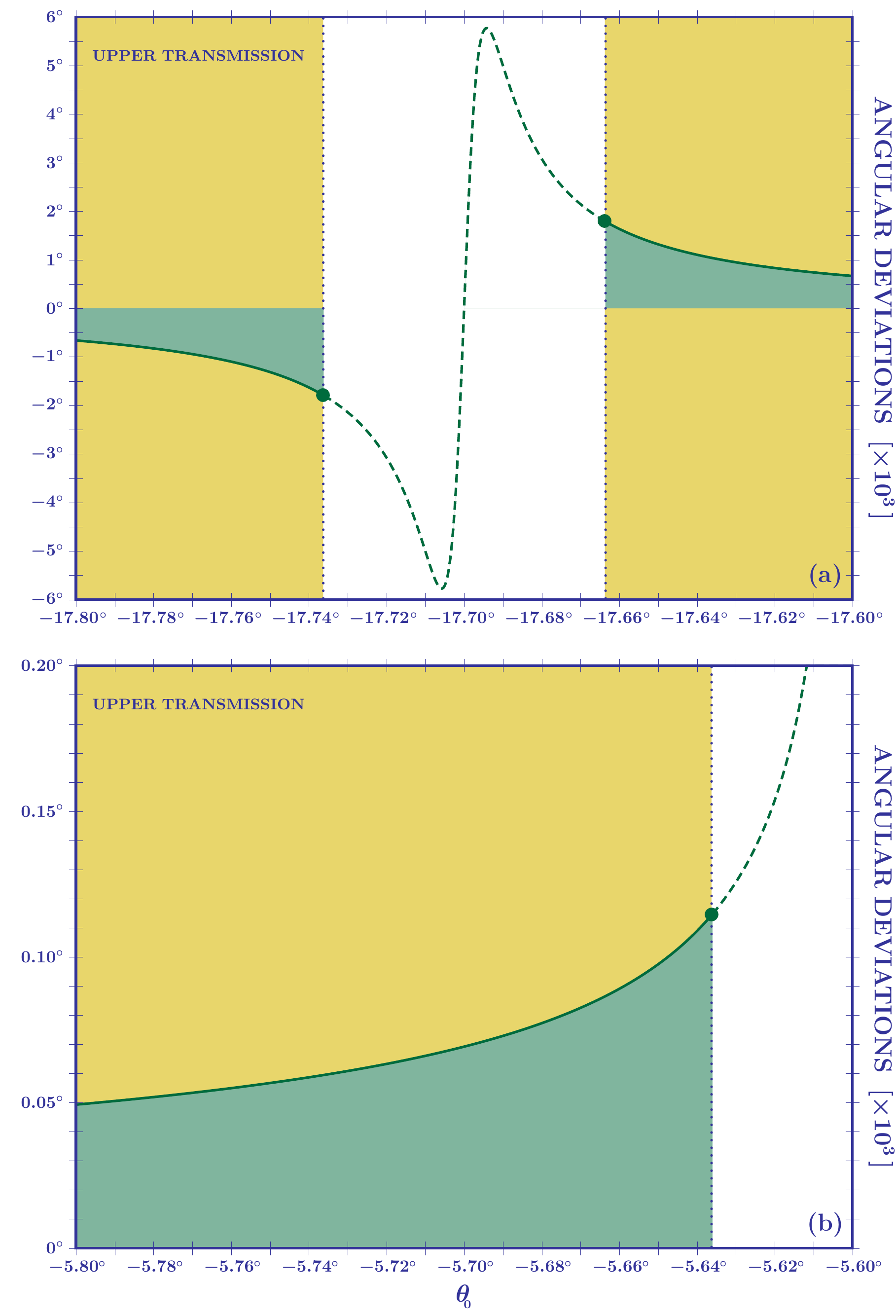}{For a BK7 prism and incident laser beam with $\lambda=0.633\,\mu{\mathrm{m}}$ and waist ${\mathrm{w}}_{\0}=1\,{\mathrm{mm}}$,
the Brewster region for internal reflection and the critical region for the upper transmission are plotted in (a) and (b), respectively.  In (a) the amplification is proportional $k\,{\mathrm{w}}_{\0}$ producing the so-called giant GH angular shift while in (b) it is proportional to
 $\sqrt{k\,{\mathrm{w}}_{\0}}$. The white background indicates the region where the concept of angular deviation is obscured due to the presence of an additional peak in the angular distribution of the reflected beam (a) or to the complex nature of the reflection coefficient (b).}
\newpage

\WideFigure{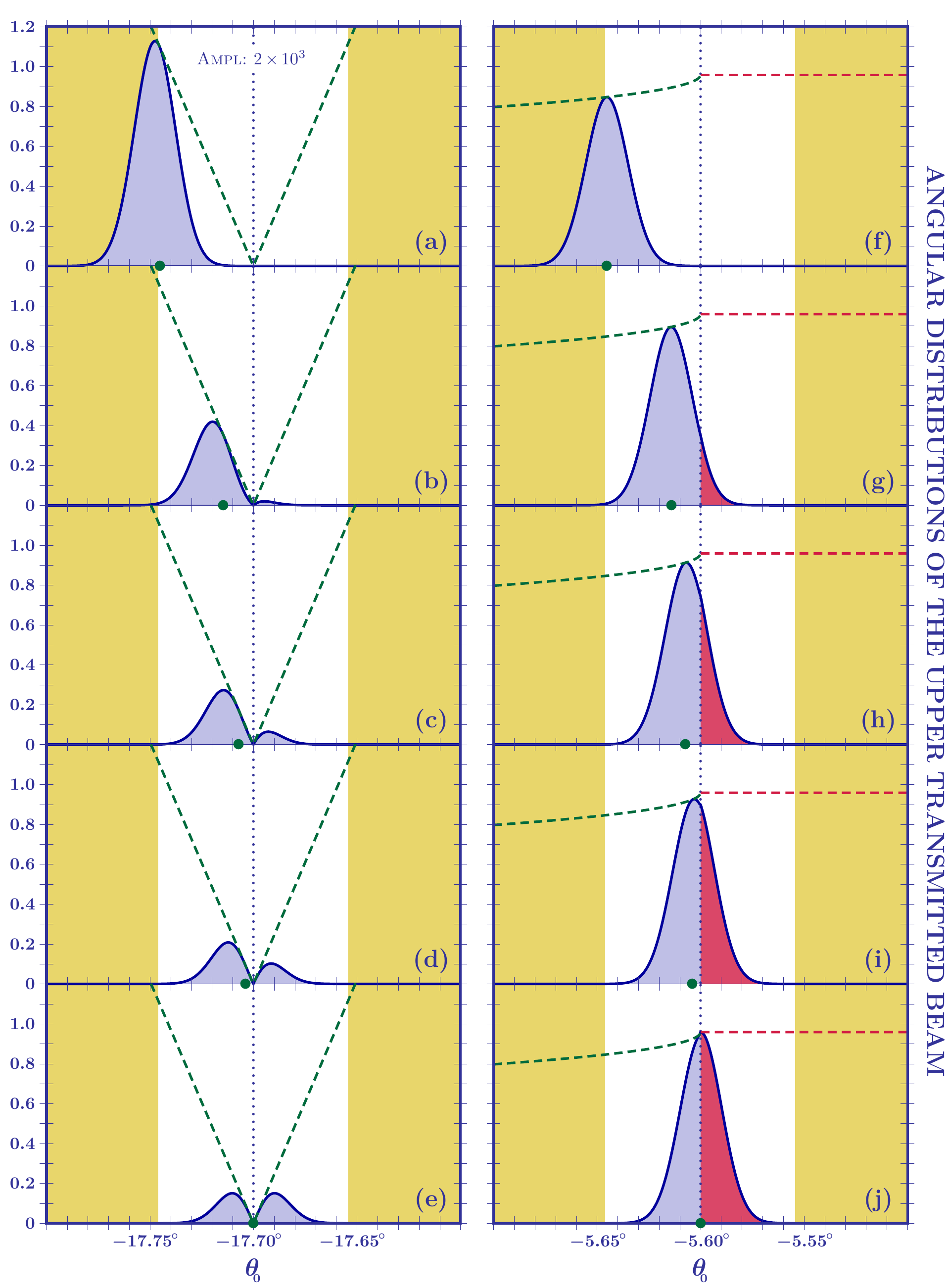}{Angular distribution absolute values of the transmitted beam in the Brewster (a-e) and critical (f-j) regions. The dashed lines represent the absolute values of the Fresnel coefficients. The concept of angular deviation works for incidence at $\theta_{_{\mathrm{B(int)}}}-\lambda/{\mathrm{w}}_{\0}$ (a) and  $\theta_{_{cri}}-\lambda/{\mathrm{w}}_{\0}$  (f). For incidence approaching the Brewster and critical angles, the presence of an additional peak, {\color{black}{see insets}} (b-e), and of a complex angular distribution, {\color{black}{red zone in}} (g-j), obscure the concept of angular deviations. }
\newpage

\WideFigure{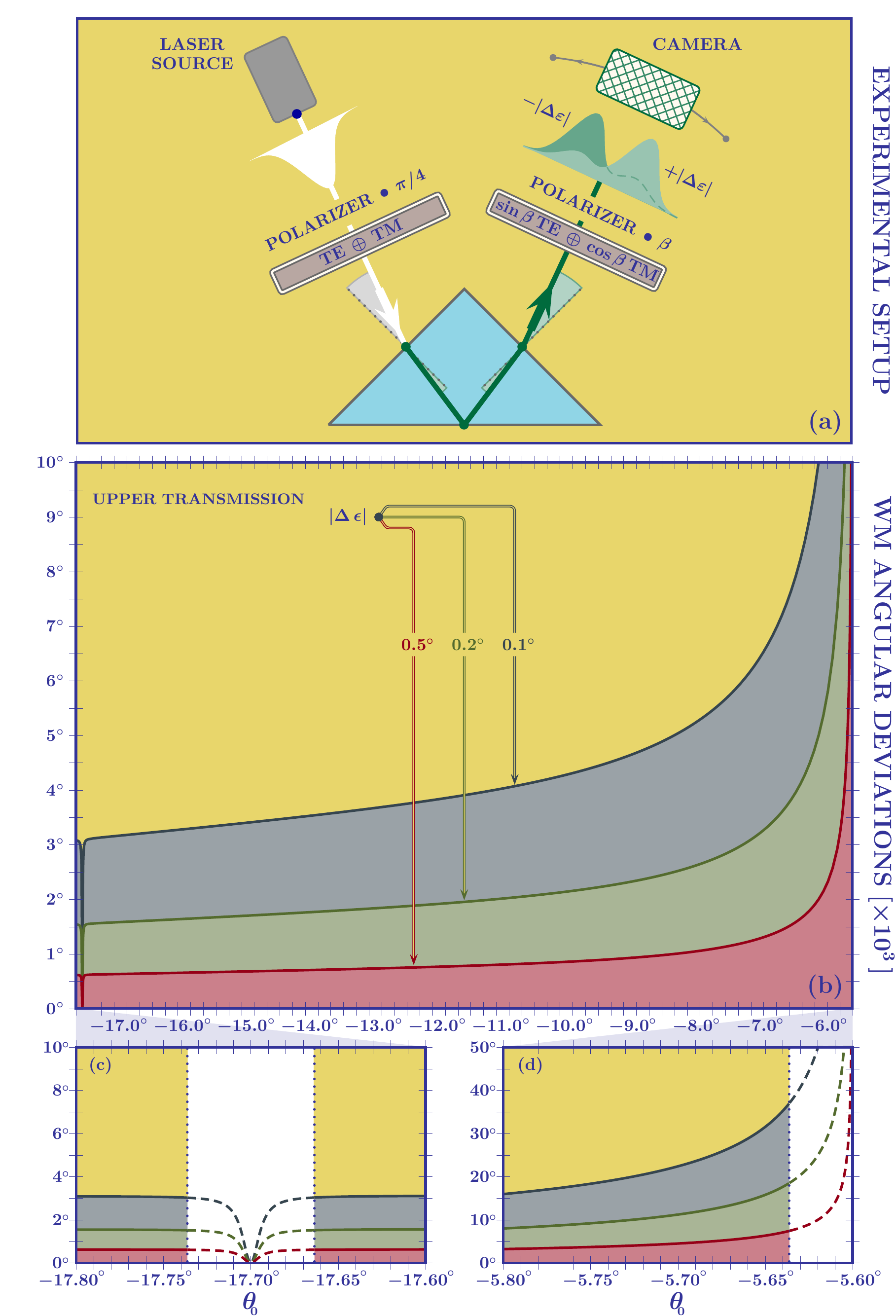}{In (a), the weak measurement (WM) experimental set-up is displayed. For angular settings of the second polarizer, $|\Delta \epsilon|=0.1^{\circ}\,,\,\,0.2^{\circ}\,,\,\,0.5^{\circ}$, we respectively find amplifications proportional to $1/|\Delta \epsilon|$, see (b). Approaching the critical angle, we find a global additional amplification. This is clearly shown in the insets (c) and (d) where we zoom to the Brewster (c) and the critical (d) regions. The breaking off of the giant GH angular shift near the Brewster region is clearly visible in (c)  and the global amplification of a factor 10 optimizing the WM analysis bear the critical angle in (d).  }


\newpage

\begin{thebibliography}{99}


\bibitem{born}
\refb{M. Born and E. Wolf}{Principles of optics}{Cambridge UP, Cambridge}{1999}

\bibitem{saleh}
\refb{B.\,E.\,A. Saleh and M.\,C. Teich}{Fundamentals of Photonics}{Wiley \& Sons, New Jersey}{2007}


\bibitem{GH1}
\refa{F. Goos and H. H\"anchen}{Ein neuer und fundamentaler Versuch zur Totalreflexion}{Ann. Phys.}{436}{333-346}{1947}

\bibitem{GH2}
\refa{K. Artmann}{Berechnung der Seitenversetzung des totalreflektierten Strahles}{Ann. Phys.}{437}{87-102}{1948}

\bibitem{GH3}
\refa{F. Goos and H. H\"anchen}{Neumessung des Strahlwersetzungseffektes bei Totalreflexion}{Ann. Phys.}{440}{251-252}{1949}

\bibitem{GH4}
\refa{B. R. Horowitz and T. Tamir}{Lateral displacement of a light beam at a dielectric interface}{J. Opt. Soc. Am.}{61}{586-594}{1971}

\bibitem{GH5}
\refa{K. Yasumoto and Y. Oishi}{A new evaluation of the Goos-H\"anchen shift and associated time delay}{J. Appl. Phys.}{54}{2170-2176}{1983}


\bibitem{GH6}
\refa{H. M. Lai, F.C. Cheng, and W.K. Tang}{Goos-H\"anchen effect around and off the critical angle}{J. Opt. Soc. Am. A}{3}{550-557}{1986}


\bibitem{GH7}
\refa{W. Nasalski, T. Tamir, L. Lin}{Displacement of the intensity peak in narrow beams reflected at a dielectric interface}{J. Opt. Soc. Am. A}{5}{132-140}{1988}


\bibitem{GH8}
\refa{S. R. Seshadri}{Goos-H\"anchen beam shift at total internal reflection}{J. Opt. Soc. Am. A}{5}{583-585}{1988}

\bibitem{GH9}
\refa{M. P. Ar\'aujo, S. A. Carvalho, and S. De Leo}{The frequency crossover for the Goos-H\"anchen shift}{J. Mod. Opt.}{60}{1772-1780}{2013}

\bibitem{GH10}
\refa{M. P. Ar\'aujo, S. De Leo, and G. G. Maia}{Closed-form expression for the Goos-Hanchen lateral displacement}{Phys. Rev.  A}{93}{023801-10}{2016}



\bibitem{GHIF0}
\refa{A. Aiello and J. P. Woerdman}{Role of beam propagation in Goos-H\"anchen and Imbert-Fedorov shifts}{Opt. Lett.}{33}{1437-1439}{2008}


\bibitem{GHIF1}
\refa{C. Prajapati and D. Ranganathan}{Goos-H\"anchen and Imbert-Federov shifts for Hermite-Gauss beams}{J. Opt. Soc. Am. A}{29}{1377-1382}{2012}

\bibitem{GHIF2}
\refa{A. Aiello}{Goos-H\"anchen and Imbert-Federov shifts: a novel perspective}{New J. Phys.}{14}{013058-12}{2012}

\bibitem{GHIF3}
\refa{K. Y. Bliokh and A. Aiello}{Goos-H\"anchen and Imbert-Fedorov beam shifts: an overview}{J. Opt.}{15}{014001-16}{2013}

\bibitem{GHIF4}
\refa{C. Prajapati and D. Ranganathan}{The effect of spectral width on Goos-H\"anchen and Imbert-Federov shifts}{J. Opt.}{15}{025703-10}{2013}


\bibitem{ANG1}
\refa{J. W. Ra, H. L. Bertoni, and L. B. Felsen}{Reflection and transmission of beams at a dielectric interface}{SIAM J. Appl. Math.}{24}{396-413}{1973}

\bibitem{ANG2}
\refa{Y. M. Antar and W. M. Boerner}{Gaussian beam interaction with a planar dielectric interface}{Can. J. Phys.}{52}{962-972}{1974}

\bibitem{ANG3}
\refa{I. A. White, A. W. Snyder, and C. Pask}{Directional change of beams undergoing partial reflection}{J. Opt. Soc. Am.}{67}{703-705}{1977}

\bibitem{ANG4}
\refa{C. C. Chan and T. Tamir}{Angular shift of a Gaussian beam reflected near the Brewster angle}{Opt. Lett.}{10}{378-380}{1985}

\bibitem{ANG5}
\refa{C. C. Chan and T. Tamir}{Beam phenomena at and near critical incidence upon a dielectric interface}{J. Opt. Soc. Am. A}{4}{655-663}{1987}

\bibitem{ANG6}
\refa{A. Aiello and J. P. Woerdman}{Theory of angular Goos-H\"anchen shift near Brewster incidence}{arXiv:0903.3730}{[physics.optics]}{1-13}{2009}

\bibitem{CrossPol}
\refa{A. Aiello, M. Merano, and J. P. Woerdman}{Brewster Cross Polarization}{Opt. Lett.}{34}{1207-1209}{2009}


\bibitem{ExpANG1}
\refa{D. M\"uller, D. Tharanga, A. A. Stahlhofen, and G. Nimtz}{Nonspecular shifts of microwaves in partial reflection}{Europhys. Lett.}{73}{526-532}{2006}

\bibitem{ExpANG2}
\refa{M. Merano, A. Aiello, M. P. van Exter, and J. P. Woerdman}{Observing angular deviations in the specular reflection of a light beam}{Nature Photonics}{3}{337-340}{2009}


\bibitem{FF}
\refa{H. E. Tureci and A. Douglas Stone}{Deviation from Snell's law for beams transmitted near the critical angle: application to microcavity lasers}{Opt. Lett.}{27}{7-9}{2002}

\bibitem{MicCav}
\refa{E. G. Altmann, G. Del Magno, and M. Hentschel}{Non-Hamiltonian dynamics in optical microcavities resulting from wave-inspired corrections to geometric optics}{EPL}{84}{10008-p6}{2008}



\bibitem{ExpFF1}
\refa{C. Gmachl, F. Capasso, E. E. Narimanov, J. U. N\"ockel, A. D. Stone, J. Faist, D. L. Sivco, and A. Y. Cho}{High-power directional emission from microlasers with chaotic resonators}{Science}{280}{1556-1564}{1998}




\bibitem{AGHE}
\refa{M. Ara\'ujo, S. A. Carvalho, and S. De Leo}{The asymmetric Goos-H\"anchen effect}{J. Opt.}{16}{015702-7}{2014}


\bibitem{Break}
\refa{M. P. Ara\'ujo, S. A. Carvalho, and S. De Leo}{Maximal breaking of symmetry at critical angle and closed-form expression for angular deviations of the Snell law}{Phys. Rev. A}{90}{033844-11}{2014}

\bibitem{Axial}
\refa{M. P. Ara\'ujo, S. De Leo and G. G. Maia}{Axial dependence of optical weak measurements in the critical region}{J. Opt.}{17}{035608-10}{2015}


\bibitem{CGH}
\refa{O. Santana, S. Carvalho, S. De Leo, and Lu\'{\i}s de Araujo}{Weak measurement of the composite Goos-H\"anchen shift in the critical region}{Opt. Lett.}{41}{3884-3887}{2016}


\bibitem{ANGFF}
\refa{J. B. G\"otte, S. Shinohara, and M. Hentschel}{Are Fresnel filtering and the angular Goos-H\"anchen shift the same?}{J. Opt.} {15}{014009-8}{2013}



\bibitem{GiantGH}
\refa{V. J. Yallapragada, A. P. Ravishankar, and G. L. Mulay}{Observation of giant Goos-H\"{a}nchen and angular shifts at designed metasurfaces}{Sci. Rep.  doi: 10.1038/srep19319}{6}{19319}{2016}










\bibitem{WM1}
\refa{Y. Aharonov, D. Z. Albert, and L. Vaidman}{How the result of a measurement of a component of the spin of a spin 1/2 particle can turn out to be 100}{Phys. Rev. Lett.}{60}{1351-1354}{1988}

\bibitem{WM2}
\refa{B. E. Y. Svensson}{Pedagogical Review of Quantum Measurement Theory with an Emphasis on Weak Measurements}{Quanta DOI: 10.12743/quanta.v2i1.12}{2(1)}{18-49}{2013}

\bibitem{WM3}
\refa{I. M. Duck and P. M. Stevenson}{The sense in which a ``weak measurement'' of a spin-$1/2$ particle's spin component yields a value 100}{Phys. Rev. D}{40}{2112-2117}{1989}


\bibitem{ExpWM1}
\refa{G. Jayaswal, G. Mistura, and M. Merano}{Weak measurement of the Goos-H\"{a}nchen shift}{Opt. Lett.}{38}{1232-1234}{2013}

\bibitem{ExpWM2}
\refa{S. Goswami, M. Pal, A. Nandi, P. K. Panigrahi, and N.Ghosh}{Simultaneous weak value amplification of angular Goos-H\"anchen and Imbert-Fedorov shifts in partial reflection}{Opt. Lett.}{39}{6229-6232}{2014}

\bibitem{ExpWM3}
\refa{G. Jayaswal, G. Mistura, M.  Merano}{Observing angular deviations in light-beam reflection via weak measurements}{Opt. Lett.}{39}{6257-6260}{2014}



\bibitem{geophs}
\refa{S. A. Carvalho and S. De Leo}{The use of the stationary phase method as a mathematical tool to determine the path of optical beams}{Am. J. Phys.}{83}{249-255}{2015}


\bibitem{profTeo}
\refa{M. P. Ara\'ujo, S. De Leo, and M. Lima}{Transversal symmetry breaking and axial spreading modification for Gaussian optical beams}{J. Mod. Opt.}{63}{417-427}{2016}

\bibitem{profExp}
\refa{S. A. Carvalho, S. De Leo, J. A. Oliveira-Huguenin, and  L. da Silva}{Experimental confirmation of the transversal symmetry breaking in laser profiles}{J. Mod. Opt.}{64}{280-287}{2016}


\end{thebibliography}
\end{document}